\documentclass[aps,prd,superscriptaddress,preprint,tightenlines,nofootinbib]{revtex4}

\usepackage{graphicx} 
\usepackage{dcolumn}  
\usepackage{bm}       
\usepackage{longtable}

\newcommand{\vcs}{V_{cs}}
\newcommand{\vcd}{V_{cd}}
\newcommand{\vcq}{V_{cd(s)}}
\newcommand{\vub}{V_{ub}}
\newcommand{\fz}{f_+(0)}
\newcommand{\chargedkenu}{D^0\to K^- e^+ \nu_e}
\newcommand{\chargedpienu}{D^0\to \pi^- e^+ \nu_e}
\newcommand{\neutralkenu}{D^+ \to \bar{K}^0 e^+ \nu_e}
\newcommand{\neutralpienu}{D^+ \to \pi^0 e^+ \nu_e}
\newcommand{\rhoenu}{D^0\to \rho^- e^+ \nu_e}
\newcommand{\kpi}{\bar{D}^0\to K^+ \pi^-}
\newcommand{\kpipiz}{\bar{D}^0\to K^+ \pi^-\pi^0}
\newcommand{\kpipipi}{\bar{D}^0\to K^+ \pi^-\pi^-\pi^+}
\newcommand{\kpipi}{D^-\rightarrow K^+\pi^-\pi^-}
\newcommand{\kpipipiz}{D^-\rightarrow K^+\pi^-\pi^-\pi^0}
\newcommand{\kzpi}{D^-\rightarrow K^0_S\pi^-}
\newcommand{\kzpipiz}{D^-\rightarrow K^0_S\pi^-\pi^0}
\newcommand{\kzpipipi}{D^-\rightarrow K^0_S\pi^-\pi^-\pi^+}
\newcommand{\kkpi}{D^-\rightarrow K^+K^-\pi^-}

\newcommand{\invpb}{{\rm pb}^{-1}}
\newcommand{\etal}{{\it et al.}}
\newcommand{\DG}{\Delta\Gamma}
\newcommand{\qsq}{q^2}
\newcommand{\qsqmax}{\infty}

\begin{document}

\preprint{CLNS 09/2049}       
\preprint{CLEO 09-02}         


\title{\boldmath Improved measurements of $D$ meson semileptonic decays to $\pi$ and $K$ mesons}



\author{D.~Besson}
\affiliation{University of Kansas, Lawrence, Kansas 66045, USA}
\author{T.~K.~Pedlar}
\author{J.~Xavier}
\affiliation{Luther College, Decorah, Iowa 52101, USA}
\author{D.~Cronin-Hennessy}
\author{K.~Y.~Gao}
\author{J.~Hietala}
\author{Y.~Kubota}
\author{T.~Klein}
\author{R.~Poling}
\author{A.~W.~Scott}
\author{P.~Zweber}
\affiliation{University of Minnesota, Minneapolis, Minnesota 55455, USA}
% ...
\collaboration{CLEO Collaboration}
\noaffiliation


\date{\today}

\begin{abstract}
Using the entire CLEO-c $\psi\left(3770\right)\rightarrow D\bar{D}$ event sample, corresponding to an integrated luminosity of $818 \nolinebreak ~\mathrm{pb}^{-1}$ and approximately 5.4 million $D\bar{D}$ events, we present a study of the decays $\chargedpienu$, $\chargedkenu$, $\neutralpienu$, and $\neutralkenu$.  Via a tagged analysis technique, in which one $D$ is fully reconstructed in a hadronic mode, partial rates for semileptonic decays by the other $D$ are measured in several $q^2$ bins.  We fit these rates using several form factor parameterizations and report the results, including form factor shape parameters and the branching fractions
${\mathcal B\left(\chargedpienu\right)}  = (0.288\pm0.008\pm0.003)$\%,
${\mathcal B \left(\chargedkenu\right)}  =  (3.50\pm0.03\pm0.04)$\%,
${\mathcal B \left(\neutralpienu\right)} =  (0.405\pm0.016\pm0.009)$\%, and
${\mathcal B \left(\neutralkenu\right)} =  (8.83\pm0.10\pm0.20)$\%,
where the first uncertainties are statistical and the second are systematic.
Taking input from lattice quantum chromodynamics (LQCD), we also find
$\left|\vcd\right| = 0.234\pm0.007\pm0.002\pm0.025$ and
$\left|\vcs\right| = 0.985\pm0.009\pm0.006\pm0.103$,
where the third uncertainties are from LQCD.

\end{abstract}

\maketitle

\section{Introduction}
Semileptonic decays are an excellent environment for precision measurements of the Cabibbo-Kobayashi-Maskawa (CKM) \cite{Kobayashi:1973fv,Cabibbo:1963yz} matrix elements.  However, because these decays are governed by both the weak and strong forces, extraction of the weak CKM parameters requires knowledge of strong interaction effects.  These can be parameterized by form factors.   Techniques such as lattice quantum chromodynamics (LQCD) offer increasingly precise calculations of these form factors, but as the uncertainties in the predictions shrink, experimental validation of the results becomes increasingly important.  Because the magnitudes of CKM matrix elements $|\vcd|$ and $|\vcs|$ are tightly constrained by CKM unitarity, semileptonic decays of $D$ mesons provide an excellent testing ground for the new theoretical predictions.  The relevance of tests using charm decays is increased by the similarity of $D$ meson decays to those of $B$ mesons, where QCD calculations are critical to extractions of $|\vub|$ in exclusive $B$ semileptonic decays \cite{Artuso:2009jw}.

We present a study of the decays $\chargedpienu$, $\chargedkenu$, $\neutralpienu$, and $\neutralkenu$ (with charged conjugate modes implied throughout this article) using 818 \nolinebreak $\invpb$ of $\psi(3770)\rightarrow D\bar{D}$ data collected by the CLEO-c detector.  Taking advantage of the fact that $D$ mesons produced near the $\psi(3770)$ resonance are produced solely as part of $D\bar{D}$ pairs, we follow a tagged technique pioneered by the Mark III Collaboration \cite{Adler:1989rw} and used in semileptonic analyses of smaller portions of CLEO-c data \cite{Ge:2008yi,Huang:2005iv,Coan:2005iu}.  Hadronically decaying $\bar{D}$ tags are first reconstructed; one then looks for the $D$ decays of interest in the remainder of each event.  This strategy suppresses backgrounds and provides an absolute normalization for branching fraction measurements.

For semileptonic decays such as those of interest here, in which  the initial and final state hadrons are pseudoscalars and the lepton mass is negligibly small, the strong interaction dynamics can be described by a single form factor $f_+\left(q^2\right)$, where $q^2$ is the invariant mass of the lepton-neutrino system.  The rate for a $D$ semileptonic decay with final state meson $P$ is given by
\begin{equation}
\frac{d\Gamma(D\rightarrow P e\nu)}{dq^2}=X\frac{G_F^2\left|\vcq\right|^2}{24\pi^3}p^3\left|f_+\left(q^2\right)\right|^2,
\label{eq:diffrate}
\end{equation}
where $G_F$ is the Fermi constant, $\vcq$ is the relevant CKM matrix element, $p$ is the momentum of the daughter meson in the rest frame of the parent $D$,  and $X$ is a multiplicative factor due to isospin, equal to 1 for all modes except $\neutralpienu$, where it is $1/2$.
The primary measurements described here are the partial decay rates $\DG=\int{\frac{d\Gamma}{dq^2}dq^2}$ in seven $q^2$ bins each for $\chargedpienu$ and $\neutralpienu$ and nine $q^2$ bins each in $\chargedkenu$ and $\neutralkenu$.  We fit the $\DG$ using several parameterizations of $f_+\left(q^2\right)$, extracting form factor shape parameters, measurements of $\left|\vcq\right|\fz$, and branching fractions.  Taking estimates of $\fz$ from theory, we also extract $|\vcd|$ and $|\vcs|$.

The remainder of this article is organized as follows:  the CLEO-c detector and event reconstruction are described in
Sec.~\ref{sec:event_recon}.  Measurements of partial rates and their systematic uncertainties are detailed in Secs.~\ref{sec:rates} and \ref{sec:systematics}, respectively.  Extractions of branching fractions, form factor shapes, and CKM parameters are reviewed in
Sec.~\ref{sec:ff_fits}, and Sec.~\ref{sec:conclusion} summarizes our results.

\section{Detector and Event Reconstruction}
\label{sec:event_recon}

The CLEO-c detector has been described in detail elsewhere \cite{Kubota:1991ww, Peterson:2002sk, artuso-2005-554}.  The 53-layer tracking system, composed of two drift chambers covering 93\% of the solid angle and enclosed by a superconducting solenoid operating with a 1-Tesla magnetic field, provides measurements of charged particle momentum with a resolution of $\sim 0.5\%$ at 700 MeV$/c$.  The tracking chambers also supply specific-ionization ($dE/dx$) information, which is combined with input from the Ring-Imaging \^{C}erenkov (RICH) detector to provide excellent discrimination between charged pions and kaons.  A 7784 crystal cesium-iodide calorimeter covering 95\% of the solid angle provides photon energy resolution of 2.2\% at $E = 1$ GeV, with a $\pi^0$ mass resolution of about 6 MeV$/c^2$, and contributes to positron identification.

The entire CLEO-c $\psi(3770)$ data sample has an integrated luminosity of 818 \nolinebreak $\invpb$, equivalent to approximately 5.4 million $D\bar{D}$ events.  The data were collected at center-of-mass energies near 3.774 GeV with a RMS spread in beam energy of approximately 2.1 MeV.  Events collected at this energy, approximately 40 MeV above the $D\bar{D}$ production threshold, are composed primarily of $D^0\bar{D}^0$, $D^+D^-$, and non-charm continuum final states.

{\sc GEANT}-based \cite{geant} Monte Carlo (MC) simulations are used to determine reconstruction efficiencies, develop line shapes for yield extraction fits, and conduct tests of the analysis procedure.  Final state radiation (FSR) is simulated using {\sc PHOTOS} \cite{Barberio:1993qi} version 2.15 with the option to simulate FSR interference enabled. A sample of generic $\psi(3770)\rightarrow D\bar{D}$ events, generated using EvtGen \cite{evtgen} and corresponding to approximately 20 times the data luminosity, was generated using input from Ref.~\cite{Yao:2006px}, combined with CLEO-c results using the initial 281 \nolinebreak $\invpb$ data sample where appropriate.  This sample, along with samples of simulated $e^+e^-\rightarrow q\bar{q}$ ($q=u$,$d$, or $s$), $e^+e^-\rightarrow\tau^+\tau^-$, and $e^+e^-\rightarrow\psi(2S)\gamma$ events corresponding to five times the data luminosity, is referred to as ``generic MC'' for the remainder of the article.  We also use a sample of $\psi(3770)\rightarrow D\bar{D}$ events in which the $D$ meson decays to one of the four studied semileptonic modes and the $\bar{D}$ decays to one of the hadronic final states used in tag reconstruction.  This sample is referred to as ``signal MC.''  In both the generic and signal MC samples, the semileptonic decays are generated using the modified pole parameterization \cite{Becirevic:1999kt} (see Sec.~\ref{sec:ff_params}) with parameters fixed to those measured in the initial 281 \nolinebreak $\invpb$ of CLEO-c data~\cite{Ge:2008yi,Cronin-Hennessey:2008}.    All simulations are corrected for small biases in the positron, charged hadron, and $\pi^0$ identification efficiencies.

 Charged pions and kaons are identified from drift chamber tracks with momentum greater than 50 MeV/$c$ and with $|\cos\theta|<0.93$, where $\theta$ is the angle between the track and the beam axis.  Charged track reconstruction efficiencies are approximately 84\% for kaons and 89\% for pions; lost tracks within $|\cos\theta|<0.93$ are almost exclusively due to particle decay in flight and material interaction in the drift chambers.  Pions and kaons are distinguished using a combination of specific ionization measurements and, if the track momentum is greater than 700 MeV/$c$, RICH detector information.   For all other tracks, hadron identity is determined using specific ionization information only.   Given a properly reconstructed track, hadron identification efficiencies are approximately 95\%, with misidentification rates of a few percent.  Identical hadron selection criteria are used in tag and semileptonic reconstructions.

Neutral pion candidates are reconstructed via $\pi^0\rightarrow\gamma\gamma$.  Photon candidates are identified from energy depositions in the calorimeter greater than 30 MeV using shower shape information.  The invariant mass of each pair of photon candidates is calculated using a kinematic fit that assumes the photons originate at the center of the detector.  This mass is required to be within three standard deviations ($3\sigma$) of the nominal $\pi^0$ mass, where $\sigma$ is determined from the kinematic fit.  The resulting $\pi^0$ energy and momentum from the fit are used in further event analysis.   Efficiencies for $\pi^0$ reconstruction vary from 40\% at a momentum of 100 MeV/$c$ to 60\% at 900 MeV/$c$.   If multiple neutral pion candidates are reconstructed opposite the tag, the candidate with the mass closest to the nominal $\pi^0$ mass is chosen.

Neutral kaon candidates are reconstructed via $K^0_S\rightarrow\pi^+\pi^-$ using vertex-constrained fits to pairs of oppositely charged intersecting tracks.  The invariant mass of the $\pi^+\pi^-$ candidate is required to be within 12 MeV/$c^2$ of the nominal $K_S^0$ mass.  This procedure results in a $K^0_S$ mass resolution of 2 -- 2.5 MeV/$c^2$ and a $K^0_S$ reconstruction efficiency of about 94\%.   If multiple $K_S^0$ candidates are reconstructed opposite a tag, the candidate with mass closest to the nominal $K^0_S$ mass is chosen.

Positron candidates are identified from tracks with momentum greater than 200 MeV/$c$ and within the solid angle $|\cos\theta|<0.9$.  Positrons are selected using a combination of specific ionization, calorimetry, and RICH detector information.  The efficiency for positron identification is about 50\% at the low momentum threshold of 200 MeV/$c$, rises sharply to 92\% at 300 MeV/$c$, and varies by a few percent as a function of momentum beyond 300 MeV/$c$.  Roughly 0.1\% of charged hadrons satisfy the positron identification criteria.  Positron momentum resolution is degraded by FSR.  We reduce this effect by identifying bremsstrahlung photon candidates in the calorimeter within $5^\circ$ of the positron candidate track and adding their 4-momenta to that of the positron candidate.  Such photons must have energy greater than 30 MeV and no associated track reconstructed in the drift chamber.

Tag candidates are reconstructed in three $\bar{D}^0$ decay modes ($\kpi$, $\kpipiz$, and $\kpipipi$) and six $D^-$ decay modes ($\kpipi$, $\kpipipiz$, $\kzpi$, $\kzpipiz$, $\kzpipipi$, and $\kkpi$).  Backgrounds are suppressed by requiring that $\Delta E=E_{\rm tag}-E_{\rm beam}$ satisfy the requirements given in Table~\ref{tab:deltae}. These cuts correspond to approximately $\pm4\sigma$, with $\sigma$ depending on the decay mode. Backgrounds are further reduced using the beam-constrained mass, $M_{\rm BC}\equiv\left({E_{\rm beam}^2/c^4-\left|{\bf P}_{\rm tag}\right|^2/c^2}\right)^{1/2}$, where $E_{\rm beam}$ is the beam energy and ${\bf P}_{\rm tag}$ is the total measured momentum of the tag candidate.  The $\bar{D}^0$ tag candidates must satisfy $1.858$ $<M_{\rm BC}<1.874$ GeV$/c^2$, while $D^-$ tag candidates are required to have $1.8628<M_{\rm BC}<1.8788$ GeV$/c^2$.  In events with multiple tag candidates, the one candidate per mode and per $D$ flavor with reconstructed energy closest to the beam energy is chosen.

\begin{table}
\caption{$\Delta E=E_{\rm tag}-E_{\rm beam}$ requirements for tag reconstruction.}
    \begin{tabular}{lc}
      \hline\hline
      Mode & Requirement (GeV) \rule[-1mm]{-1mm}{4.3mm} \\ \hline
      $\kpi$ & $|\Delta E|<0.030$ \rule[-1mm]{-1mm}{4.3mm}\\
      $\kpipiz$ &  $-0.050<\Delta E<0.044$ \\
      $\kpipipi$ & $|\Delta E|<0.020$ \\
      $\kpipi$ & $|\Delta E|<0.0232$ \\
      $\kpipipiz$ & $|\Delta E|<0.0276$ \\
      $\kzpi$ & $|\Delta E|<0.0272$ \\
      $\kzpipiz$ & $|\Delta E|<0.0366$ \\
      $\kzpipipi$ & $|\Delta E|<0.0159$ \\
      $\kkpi$ & $|\Delta E| < 0.0138$ \\
      \hline\hline
    \end{tabular}
  \label{tab:deltae}
\end{table}

Semileptonic candidates are formed from positron and hadron candidate pairs.   Although the semileptonic neutrino daughter is not detected, its energy and momentum can be inferred from the missing energy $E_{\rm miss}$ and momentum ${\bf P}_{\rm miss}$ of the event:
\begin{equation}
E_{\rm miss} = E_{\rm beam}-E_{he}
\label{eq:emiss}
\end{equation}
and
\begin{equation}
{\bf P}_{\rm miss} = -{\bf P}'_{\rm tag}-{\bf P}_{he},
\label{eq:pmiss}
\end{equation}
where the energy $E_{he}$ and momentum ${\bf P}_{he}$ of the hadron-positron system are constructed from the measured energy and momenta of the hadron, positron and any bremsstrahlung photon candidates.  The tag momentum ${\bf P}'_{\rm tag}$ is formed from the measured tag momentum with the magnitude constrained using the beam energy and $D$ mass:
${\bf P}'_{\rm tag}=[(E_{\rm beam}/c)^2-(c m_D)^2]^{1/2}\hat{\bf P}_{\rm tag}$.  All momentum vectors are boosted to the center-of-mass frame by correcting for the small $e^+e^-$ net momentum due to the beam crossing angle~($\sim 3$~mrad).

Semileptonic backgrounds are reduced by requiring that the variable $U$, defined as
\begin{equation}
U\equiv E_{\rm miss}-c|{\bf P}_{\rm miss}|,
\label{eq:udef}
\end{equation}
 satisfy $-0.10<U<0.24$ GeV for each candidate.  Additionally, the positron and hadron from the semileptonic decay are required to have opposite charge in $\chargedpienu$ and $\chargedkenu$ candidates, while the positron and tag are required to have opposite charge in $\neutralpienu$ and $\neutralkenu$.

Semileptonic candidates are partitioned into several $q^2$ bins, where the reconstructed $q^2$ is determined based on measurements of the positron and neutrino:
\begin{equation}
q^2 = \frac{1}{c^4}\left(E_\nu+E_e\right)^2-\frac{1}{c^2}\left|{\bf P}_\nu+{\bf P}_e\right|^2.
\end{equation}
The neutrino energy is taken to be the missing energy of the event, while the neutrino momentum is equated with the missing momentum scaled so that $\left|{\bf P}_{\rm miss}\right| = E_{\rm miss}$.  Because $E_{\rm miss}$ does not require measurements of the tag decay, it is a better measured quantity than $\left|{\bf P}_{\rm miss}\right|$; constraining the neutrino momentum in this manner thus improves the resolution in $q^2$.   $\chargedpienu$ and $\neutralpienu$ candidates are divided into seven $q^2$ bins, with boundaries defined by [0,\,0.3), [0.3,\,0.6), [0.6,\,0.9), [0.9,\,1.2), [1.2,\,1.5), [1.5,\,2.0), and [2.0,\,$\infty$) GeV$^2/c^4$.  In the $\chargedkenu$ and $\neutralkenu$ modes, nine $q^2$ bins are used, defined by [0,\,0.2), [0.2,\,0.4), [0.4,\,0.6), [0.6,\,0.8), [0.8,\,1.0), [1.0,\,1.2), [1.2,\,1.4), [1.4,\,1.6), and [1.6,\,$\infty$) GeV$^2/c^4$.

\section{Extraction of Partial Rates}
\label{sec:rates}

In order to measure the partial rates we first determine the number of observed tags $N^{\rm obs,\alpha}_{\rm tag}$ (``tag yields'') in each tag mode $\alpha$.   This is related to the number of tags produced in mode $\alpha$, $N^{\alpha}_{\rm tag}$, by
\begin{equation}
N^{\alpha}_{\rm tag} = \frac{N^{\rm obs,\alpha}_{\rm tag}}{\epsilon_{\rm tag}^\alpha},
\label{eq:n_tag}
\end{equation}
where $\epsilon_{\rm tag}^\alpha$ is the reconstruction efficiency for tag mode $\alpha$.  We then determine the number of events with both a tag and semileptonic candidate. These ``signal yields'' $n_j^{\rm obs,\alpha}$ are determined separately for each tag mode $\alpha$ and $q^2$ bin $j$.  The signal yields are related to the number of tag-semileptonic combinations produced in each $q^2$ bin, $n_i^\alpha$, by
\begin{equation}
n_j^{\rm obs,\alpha} = \sum_i{n_i^\alpha\epsilon_{ji}^\alpha},
\label{eq:sigeffdef}
\end{equation}
where $\epsilon^\alpha_{ij}$ are the elements of a matrix that describes the efficiency and smearing across $q^2$ bins associated with tag and semileptonic reconstruction.  As the number of tag-semileptonic combinations produced is a function of the number of tag decays and the differential semileptonic decay rate, $d\Gamma/dq^2$, we can rewrite Eq.~(\ref{eq:sigeffdef}) as
\begin{equation}
n_{j}^{\rm obs,\alpha} = N^\alpha_{\rm tag}\tau_D\sum_i{\epsilon_{ji}^\alpha\int_i{\frac{d\Gamma}{dq^2}dq^2}},
\label{eq:n_obs_1}
\end{equation}
where $\tau_D$ is the $D$ lifetime and the integration is over the width of $q^2$ bin $i$. Combining the above equations and solving for the differential rate, we obtain a simple formula for extracting the partial rates:
\begin{eqnarray}
\Delta\Gamma_i & \equiv & \int_{i}\frac{d\Gamma}{dq^2}dq^2  \nonumber \\
& = & \frac{1}{\tau_D}\frac{\epsilon_{\rm tag}^\alpha}{N^{\rm obs,\alpha}_{\rm tag}}\sum_j{\left(\epsilon^{-1}\right)^\alpha_{ij}n_{j}^{\rm obs,\alpha}}.
\label{eq:dgdef}
\end{eqnarray}
The small correlation between signal and tag yields that arises from the signal yields being a subset of the tag yields is neglected.

The following sections describe the extraction of the tag yields $N^{{\rm obs},\alpha}_{\rm tag}$, tagging efficiencies $\epsilon_{\rm tag}^\alpha$,  signal yields $n^{\rm obs,\alpha}_i$, and signal smearing and efficiency matrices $\epsilon^\alpha_{ij}$.  With all of these numbers in hand, we then extract the partial rates.

\subsection{Tag Yields and Efficiencies}
\begin{figure*}[bptb]
  \includegraphics*[width=6.0in]{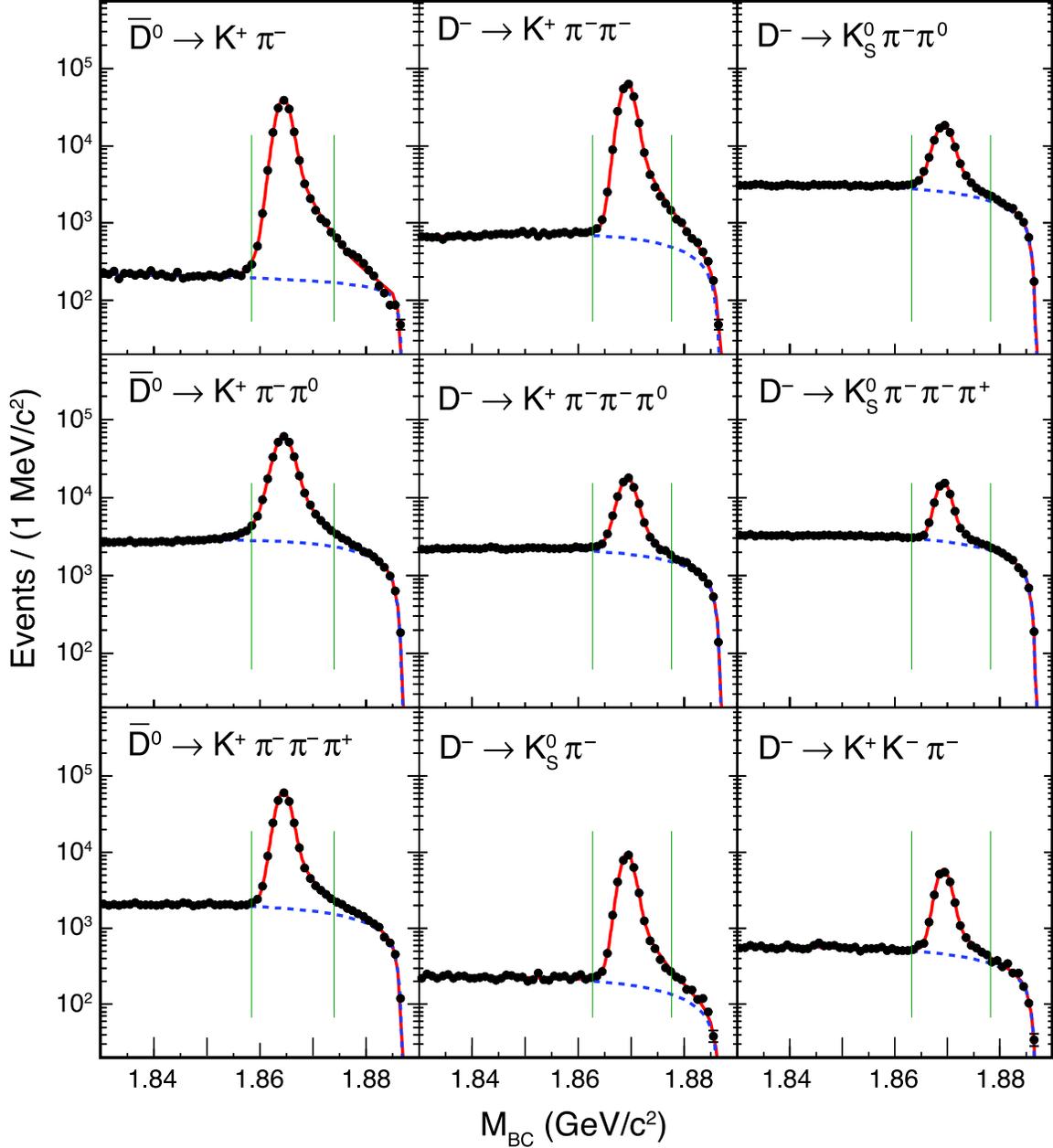}
  \caption{$M_{\rm BC}$ distributions in data (points), with fits (solid lines) and background contributions to fits (dotted lines). The vertical lines show the limits of the $M_{\rm BC}$ signal regions.}
  \label{fig:mbc_log}
\end{figure*}

The tag yields $N^{\rm obs,\alpha}_{\rm tag}$ are obtained separately for each mode $\alpha$ by performing fits to beam-constrained mass distributions.  The fitting procedure has been described in detail in \cite{:2007zt} and involves an unbinned likelihood maximization.  True tag decays are modeled using a function specially designed to take into account the natural $\psi(3770)$ line shape, beam energy resolution, momentum resolution, and initial state radiation (ISR) effects.  Tag backgrounds are modeled using an ARGUS function \cite{Albrecht:1990am}, modified so that the power parameter is allowed to float~\cite{:2007zt}.
The fits, shown in Fig.~\ref{fig:mbc_log}, are performed over a wide $1.83<M_{\rm BC}<1.89$ GeV$/c^2$ window. Tag yields, given in Table~\ref{tab:tagyields}, are obtained by subtracting the backgrounds estimated by the fits from event counts in data inside the narrower $M_{\rm BC}$ signal regions.  Also shown in Table~\ref{tab:tagyields} are tagging efficiencies, which are obtained by fitting generic MC $M_{\rm BC}$ distributions with the same procedure used in data.

\begin{table}
\caption{Tag yields and statistical uncertainties in data and tag reconstruction efficiencies. }
    \begin{tabular}{lcc}
      \hline\hline
      Mode & Yield & Efficiency(\%) \rule[-1mm]{-1mm}{4.3mm}\\ \hline
      $\kpi$ & $149616\pm392$ & 65.32 \rule[-1mm]{-1mm}{4.3mm} \\
      $\kpipiz$ & $284617\pm589$ & 35.15 \\
      $\kpipipi$ & $227536\pm517$ & 45.55 \\
      $\kpipi$ & $233670\pm497$ & 55.42 \\
      $\kpipipiz$ & $69798\pm330$ & 27.39 \\
      $\kzpi$ & $33870\pm194$ & 51.10  \\
      $\kzpipiz$ & $74842\pm357$ & 28.74 \\
      $\kzpipipi$ & $49117\pm323$ & 43.58 \\
      $\kkpi$ & $19926\pm171$ & 42.07\\
      \hline\hline
    \end{tabular}
  \label{tab:tagyields}
\end{table}

\subsection{Signal Yields and Efficiencies}
\label{sec:signal_yields}
\begin{figure*}[bptb]
  \includegraphics*[width=6in]{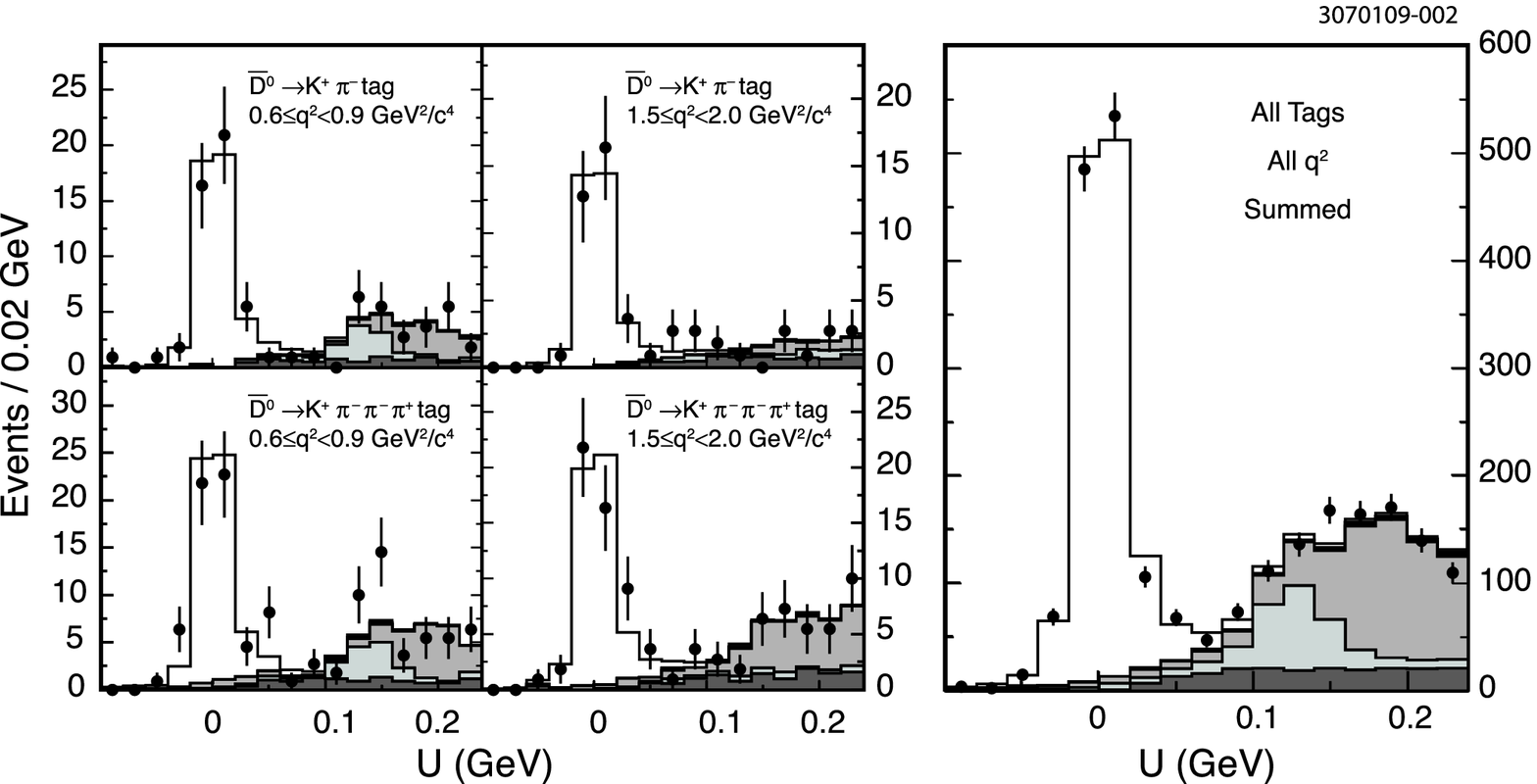}
  \caption{$U$ distributions in data (points) for $\chargedpienu$, with fit results (histograms) showing signal (clear) and background components:  $\rhoenu$ (darkest gray), $\chargedkenu$ (lightest gray), other $D\bar{D}$ (medium gray), and non-$D\bar{D}$ (black).  }
  \label{fig:u_pi}
\end{figure*}

\begin{figure*}[bptb]
  \includegraphics*[width=6in]{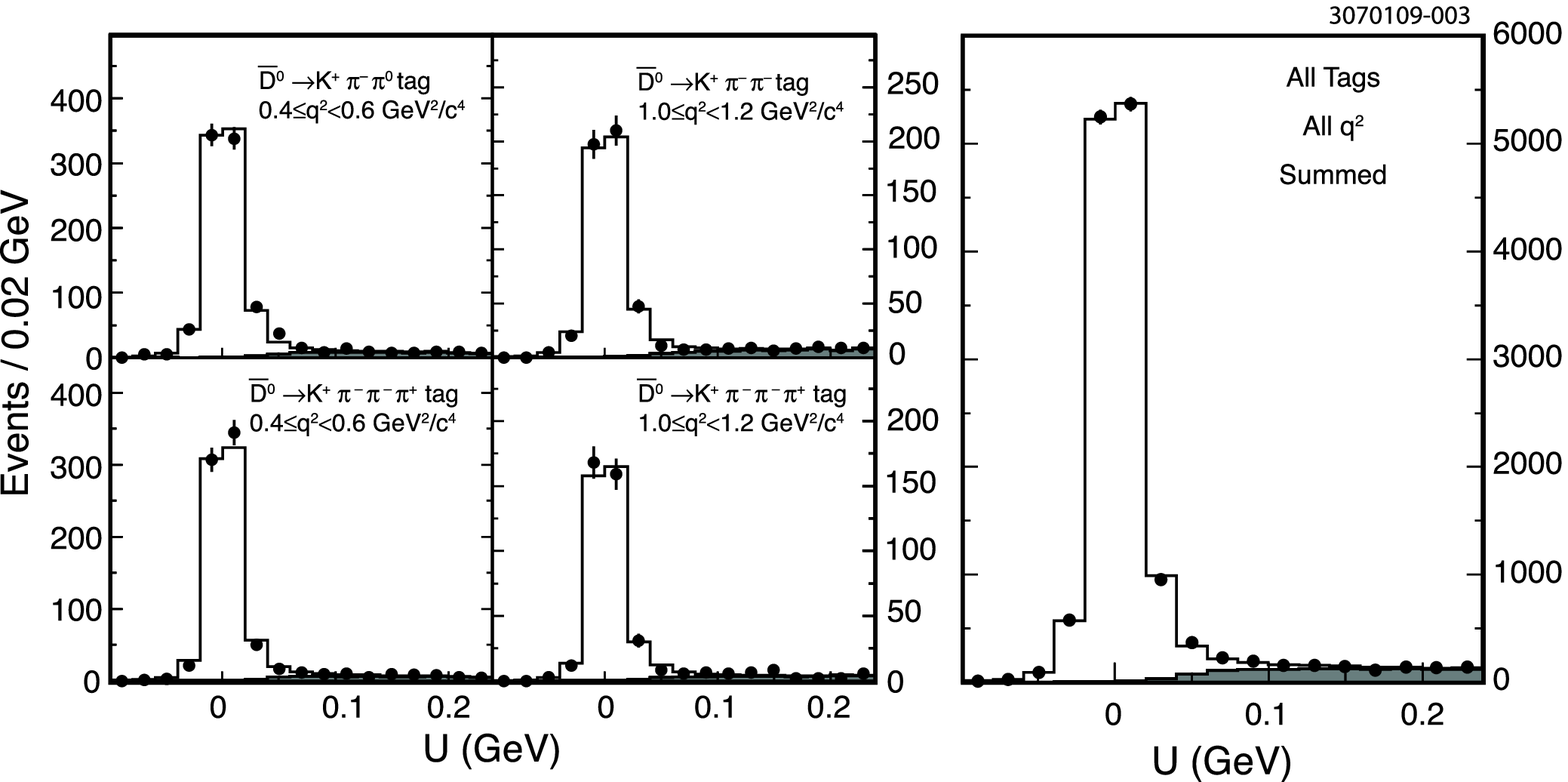}
  \caption{$U$ distributions in data (points) for $\chargedkenu$, with fit results (histograms) showing signal (clear) and background components:  $D\bar{D}$ (gray) and non-$D\bar{D}$ (black).  }
  \label{fig:u_k}
\end{figure*}

\begin{figure*}[bptb]
  \includegraphics*[width=6in]{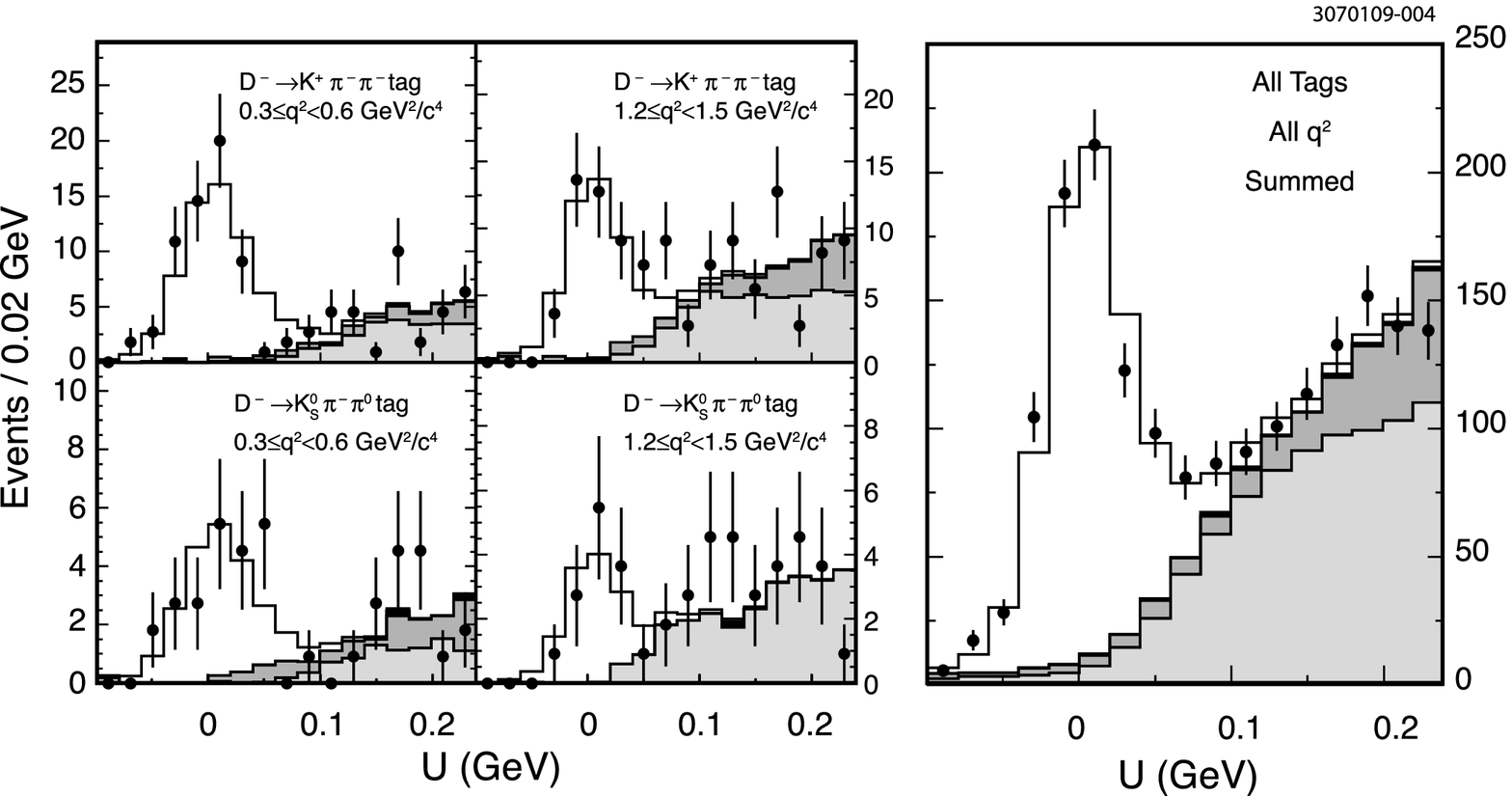}
  \caption{$U$ distributions in data (points) for $\neutralpienu$, with fit results (histograms) showing signal (clear) and background components: $\neutralkenu$ (light gray), other $D\bar{D}$ (dark gray), and non-$D\bar{D}$ (black).  }
  \label{fig:u_pi0}
\end{figure*}

\begin{figure*}[bptb]
  \includegraphics*[width=6in]{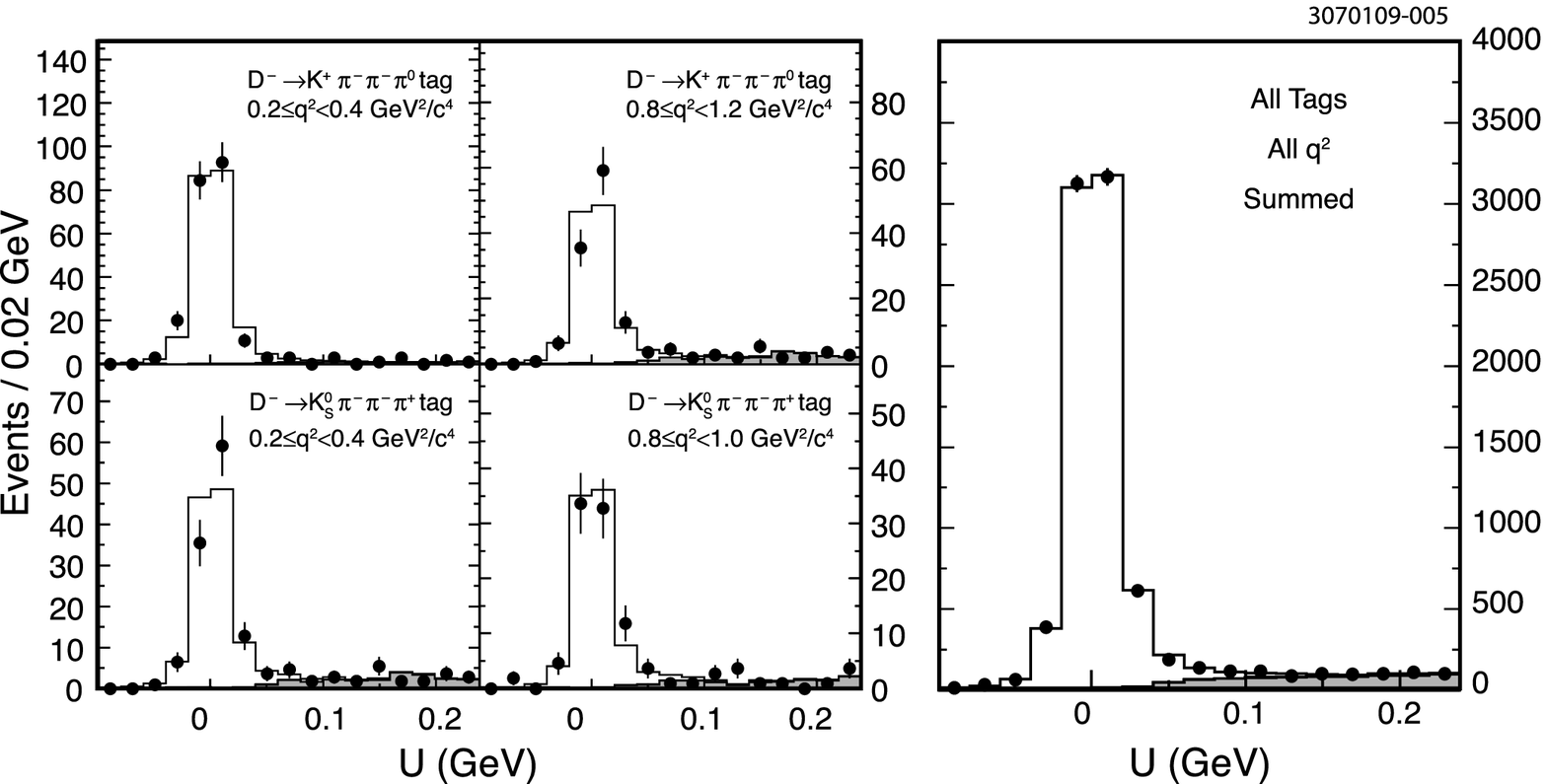}
  \caption{$U$ distributions in data (points) for $\neutralkenu$, with fit results (histograms) showing signal (clear) and background components: $D\bar{D}$ (gray) and non-$D\bar{D}$ (black).  }
  \label{fig:u_k0}
\end{figure*}

Signal yields are extracted from distributions of $U$, defined in Eq.~(\ref{eq:udef}).  Events in which both the tag and semileptonic decay have been correctly reconstructed, leaving only an undetected neutrino, are expected to peak at $U=0$, with the shape of the distribution being approximately Gaussian due to detector resolution.   Misreconstructed events and background modes generally have non-zero $U$ values.   Properly reconstructed decays are separated from backgrounds using an unbinned maximum-likelihood fit, executed independently for each semileptonic mode, each tag mode, and each $q^2$ bin. A sample of the $U$ distributions is shown in Figs.~\ref{fig:u_pi} -- \ref{fig:u_k0}.

For the fit, the $U$ distribution of signal candidates is taken from signal MC samples.  While the $U$ resolution in data is approximately 12 MeV for the modes with only charged tracks in the final state ($\chargedpienu$, $\chargedkenu$, and $\neutralkenu$) and 25 MeV for $\neutralpienu$, the distributions are slightly narrower in MC simulation.  To accommodate the poorer $U$ resolutions in data, the MC distributions are convolved with a double Gaussian with parameters fixed for each semileptonic mode to the values that maximize the fit likelihoods summed over all $q^2$ bins and tag modes.  The smearing functions are dominated by central Gaussians with widths of approximately 6 MeV in $\chargedkenu$, $\chargedpienu$, and $\neutralkenu$ and 13 MeV in $\neutralpienu$. The secondary Gaussians have normalizations of 3\% -- 7\% of the central Gaussian and have widths of 30 -- 35 MeV.  The overall normalization of the corrected signal distribution is allowed to float in each fit.

The background distributions used in the fits are taken from generic MC samples.  Backgrounds arise from both non-$D\bar{D}$ and $D\bar{D}$ events.
The $\chargedpienu$ and $\neutralpienu$ modes are subject to large backgrounds from $\chargedkenu$ and $\neutralkenu$, respectively.  To allow for variations in $\pi^-$ and $\pi^0$ fake rates between the data and MC simulations, the normalizations of these components are fixed to the values that minimize the fit likelihood summed over all $q^2$ bins and tag modes.
The $\chargedpienu$ mode is also subject to a large background from $\rhoenu$ events; the normalization of this background is fixed using the known branching fraction and the tag yields in data and MC samples.  The remaining $D\bar{D}$ backgrounds occur due to misreconstruction of either the semileptonic or tag decay, although the largest backgrounds are composed of events with a correctly reconstructed tag but misreconstructed semileptonic decay.   In each semileptonic mode, all of the $D\bar{D}$ backgrounds not discussed above are combined into a single background distribution with the normalization allowed to float in each fit.
The normalization of the non-$D\bar{D}$ distribution is fixed using the ratio of luminosities in data and MC samples.

Because each $q^2$ bin and tag mode is treated separately, the total numbers of fits for $\chargedpienu$, $\chargedkenu$, $\neutralpienu$, and $\neutralkenu$  are 21, 27, 42, and 54, respectively.  Figs.~\ref{fig:u_pi} -- \ref{fig:u_k0} show four individual fits, as well as the summed fit results for all $q^2$ bins and tag modes.


\begingroup
\squeezetable
\begin{table*}
\caption{Selected efficiency matrices $\epsilon_{ij}$ in percent for $\chargedpienu$ and $\chargedkenu$.  The column gives the true $q^2$ bin $j$, while the row gives the reconstructed (``Rec'') $q^2$ bin $i$.  The elements account for the reconstruction efficiencies of both the tag and the semileptonic decay.  The statistical uncertainties in the least significant digits are given in the parentheses.}
    \begin{tabular}{cccccccccc}
      \hline\hline
      \multicolumn{10}{c}{\boldmath ${\chargedpienu, \kpi}$}\\
      Rec $q^2$  &  \multicolumn{9}{l}{True $q^2$ (GeV$^2/c^4$) }\\
     (GeV$^2/c^4$)         & [0,0.3) &[0.3,0.6)&[0.6,0.9)&[0.9,1.2)&[1.2,1.5)&[1.5,2.0)&[2.0,$\qsqmax$) & &\\ \hline
$[0,\,0.3)$     &41.25(34) &1.19(8)  &0.02(1)  &0.00(0)  &0.00(0)  &0.00(0) &0.00(0) & &\\
$[0.3,\,0.6)$   &0.76(6)   &42.57(36)&1.55(10) &0.01(1)  &0.00(0)  &0.00(0) &0.00(0) & &\\
$[0.6,\,0.9)$   &0.04(1)   &1.12(8)  &44.65(38)&1.54(10) &0.02(1)  &0.00(0) &0.00(0) & &\\
$[0.9,\,1.2)$   &0.02(1)   &0.08(2)  &1.09(8)  &44.77(41)&1.37(10) &0.03(1) &0.00(0) & &\\
$[1.2,\,1.5)$   &0.01(1)   &0.03(1)  &0.09(2)  &1.33(9)  &46.11(44)&0.91(8) &0.00(0) & &\\
$[1.5,\,2.0)$   &0.01(1)   &0.02(1)  &0.02(1)  &0.11(3)  &1.20(10) &47.01(40)&0.74(8) & &\\
$[2.0,\,\qsqmax)$&0.00(0)   &0.00(0)  &0.01(1)  &0.02(1)  &0.04(2)  &0.56(6) &47.28(48)\\
      \hline
      \multicolumn{10}{c}{\boldmath ${\chargedkenu, \kpipiz}$} \rule[-1mm]{-1mm}{4.3mm}\\
       Rec $q^2$&  \multicolumn{9}{l}{True $q^2$ (GeV$^2/c^4$) }\\
     (GeV$^2/c^4$)         & [0,\,0.2) &[0.2,\,0.4)&[0.4,\,0.6)&[0.6,\,0.8)&[0.8,\,1.0)&[1.0,\,1.2)&[1.2,\,1.4)&[1.4,\,1.6)&[1.6,\,$\qsqmax$)\\ \hline
$[0,\,0.2)$     & 19.70(4) & 0.80(1) & 0.03(0) & 0.00(0) & 0.00(0) & 0.00(0) & 0.00(0) & 0.00(0) & 0.00(0) \\
$[0.2,\,0.4)$   & 0.45(1)  & 19.80(5)& 1.03(1) & 0.03(0) & 0.00(0) & 0.00(0) & 0.00(0) & 0.00(0) & 0.00(0) \\
$[0.4,\,0.6)$   & 0.01(0)  & 0.54(1) & 20.58(5)& 1.12(1) & 0.03(0) & 0.01(0) & 0.00(0)& 0.00(0) & 0.00(0)\\
$[0.6,\,0.8)$   & 0.01(0)  & 0.02(0) & 0.61(1) & 21.32(6)& 1.12(2) & 0.03(0) & 0.01(0) & 0.00(0) & 0.01(0)\\
$[0.8,\,1.0)$   & 0.01(0)  & 0.01(0) & 0.03(0) & 0.63(1) & 21.92(6)& 1.03(2) & 0.01(0) & 0.00(0) & 0.00(0)\\
$[1.0,\,1.2)$   & 0.00(0)  & 0.01(0) & 0.02(0) & 0.03(0) & 0.59(1) & 21.64(7)& 0.95(2) & 0.01(0) & 0.00(0)\\
$[1.2,\,1.4)$   & 0.00(0)  & 0.00(0) & 0.01(0) & 0.01(0) & 0.01(0) & 0.51(1) & 21.08(9)& 0.88(3) & 0.01(0)\\
$[1.4,\,1.6)$   & 0.00(0)  & 0.00(0) & 0.00(0) & 0.00(0) & 0.00(0) & 0.01(0) & 0.39(1) & 20.05(11)& 0.79(4)\\
$[1.6,\,\qsqmax)$& 0.00(0)  & 0.00(0) & 0.00(0) & 0.00(0) & 0.00(0) & 0.00(0) & 0.00(0) & 0.25(1) & 16.72(17)\\
\hline\hline
    \end{tabular}
  \label{tab:eff1}
\end{table*}
\endgroup

\begingroup
\squeezetable
\begin{table*}
\caption{Selected efficiency matrices $\epsilon_{ij}$ in percent for $\neutralpienu$ and $\neutralkenu$.  The column gives the true $q^2$ bin $j$, while the row gives the reconstructed (``Rec'') $q^2$ bin $i$.  The elements account for the reconstruction efficiencies of both the tag and the semileptonic decay.  The statistical uncertainties in the least significant digits are given in the parentheses.}
    \begin{tabular}{cccccccccc}
     \hline\hline
      \multicolumn{10}{c}{\boldmath ${\neutralpienu, \kpipi}$} \rule[-1mm]{-1mm}{4.3mm}\\
      Rec $q^2$&  \multicolumn{9}{l}{True $q^2$ (GeV$^2/c^4$) }\\
      (GeV$^2/c^4$)         & [0,\,0.3) &[0.3,\,0.6)&[0.6,\,0.9)&[0.9,\,1.2)&[1.2,\,1.5)&[1.5,\,2.0)&[2.0,\,$\qsqmax$)\\ \hline
$[0,\,0.3)$     &22.44(20) &0.83(5) &0.02(1) &0.00(0) &0.00(0) &0.00(0) &0.00(0)\\
$[0.3,\,0.6)$   &1.23(5)  &21.69(21)&1.02(5) &0.01(1) &0.00(0) &0.00(0) &0.00(0)\\
$[0.6,\,0.9)$   &0.03(1)  &1.62(6) &21.23(22)&1.14(6) &0.01(1) &0.00(0) &0.00(0)\\
$[0.9,\,1.2)$   &0.02(1)  &0.03(1) &1.75(7) &21.12(23)&1.05(6) &0.00(0) &0.00(0)\\
$[1.2,\,1.5)$   &0.02(1)  &0.03(1) &0.06(1) &1.61(7) &19.72(25)&0.65(4) &0.00(0)\\
$[1.5,\,2.0)$   &0.02(1)  &0.03(1) &0.04(1) &0.13(2) &1.47(7) &20.50(22)&0.49(5)\\
$[2.0,\,\qsqmax)$&0.17(2)  &0.19(2) &0.31(3) &0.47(4) &0.70(5) &1.65(7) &22.81(27)\\
      \hline
      \multicolumn{10}{c}{\boldmath ${\neutralkenu, \kpipipiz}$} \rule[-1mm]{-1mm}{4.3mm}\\
      Rec $q^2$&  \multicolumn{9}{l}{True $q^2$ (GeV$^2/c^4$) }\\
      (GeV$^2/c^4$)         & [0,\,0.2) &[0.2,\,0.4)&[0.4,\,0.6)&[0.6,\,0.8)&[0.8,\,1.0)&[1.0,\,1.2)&[1.2,\,1.4)&[1.4,\,1.6)&[1.6,\,$\qsqmax$)\\ \hline
$[0,\,0.2)$     & 5.06(3)  & 0.21(1) & 0.00(0) & 0.00(0) & 0.00(0) & 0.00(0) & 0.00(0) & 0.00(0) & 0.00(0) \\
$[0.2,\,0.4)$   & 0.11(0)  & 4.98(3) & 0.24(1) & 0.00(0) & 0.00(0) & 0.00(0) & 0.00(0) & 0.00(0) & 0.00(0) \\
$[0.4,\,0.6)$   & 0.00(0)  & 0.15(1) & 5.09(3) & 0.25(1) & 0.01(0) & 0.00(0) & 0.00(0) & 0.00(0) & 0.00(0)\\
$[0.6,\,0.8)$   & 0.00(0)  & 0.00(0) & 0.16(1) & 5.12(3)& 0.28(1) & 0.00(0) & 0.00(0) & 0.00(0) & 0.00(0)\\
$[0.8,\,1.0)$   & 0.00(0)  & 0.00(0) & 0.00(0) & 0.15(1) & 5.13(4)& 0.26(1) & 0.00(0) & 0.00(0) & 0.00(0)\\
$[1.0,\,1.2)$   & 0.00(0)  & 0.00(0) & 0.00(0) & 0.01(0) & 0.15(1) & 5.14(4)& 0.24(1) & 0.00(0) & 0.00(0)\\
$[1.2,\,1.4)$   & 0.00(0)  & 0.00(0) & 0.00(0) & 0.00(0) & 0.01(0) & 0.12(1) & 5.29(5)& 0.22(1) & 0.01(0)\\
$[1.4,\,1.6)$   & 0.00(0)  & 0.00(0) & 0.00(0) & 0.00(0) & 0.00(0) & 0.01(0) & 0.12(1) & 5.34(7)& 0.26(3)\\
$[1.6,\,\qsqmax)$& 0.00(0)  & 0.00(0) & 0.01(0) & 0.01(0) & 0.01(0) & 0.01(0) & 0.01(0) & 0.08(1) & 5.45(11)\\ \hline\hline
    \end{tabular}
  \label{tab:eff2}
\end{table*}
\endgroup

The signal efficiency matrix elements $\epsilon_{ij}^\alpha$, as defined in Eq.~(\ref{eq:sigeffdef}), are obtained from signal MC simulations and corrected for previously determined deviations from positron, charged hadron, and $\pi^0$ identification efficiencies in data.  Each $\epsilon_{ij}^\alpha$ gives the fraction of
events generated in $q^2$ bin $j$ with tag mode $\alpha$ that are
reconstructed in $q^2$ bin $i$ with the same tag.  The efficiency
matrix thus accounts for reconstruction of both the signal and tag
decays.  The $\neutralpienu$ efficiencies include the $\pi^0\rightarrow\gamma\gamma$ branching fraction~\cite{Amsler:2008zz} and $\neutralkenu$ efficiencies include the $K^0_S$ fraction of the $K^0$ and $K^0_S\rightarrow \pi^+\pi^-$ branching fraction~\cite{Amsler:2008zz}.  In total, there are eighteen efficiency matrices -- one for each tag and semileptonic mode combination.  Tables~\ref{tab:eff1} and \ref{tab:eff2} provide four examples of these
matrices.
  The diagonal elements, giving the efficiency for the tag and semileptonic decays to be reconstructed in the correct $q^2$ bin, vary from
5\% -- 50\% depending on semileptonic mode, tag mode, and $q^2$.  The neighboring off-diagonal elements, giving the efficiencies for the tag and semileptonic decay to be reconstructed in the wrong $q^2$ bin, range between 1\% and 10\% of the diagonal elements.  The signal yields summed over tag modes both before and after correction by these matrices are shown in Table~\ref{tab:sigyields}.

\begingroup
\squeezetable
\begin{table*}
\caption{Signal yields, both raw ($n_i^{\rm{obs},\alpha}$) and corrected for $q^2$ smearing and reconstruction efficiency ($n_i^\alpha$), and partial rates ($\Delta\Gamma_i$).  Statistical uncertainties in the least significant digits are given in parentheses.}
  \label{tab:sigyields}

    \begin{tabular}{cccccccccc} 
\hline \hline 
& \multicolumn{9}{c}{$\mathbf{\chargedpienu}$} \\

$q^2$ (GeV$^2/c^4$) & [0,\,0.3) &[0.3,\,0.6)&[0.6,\,0.9)&[0.9,\,1.2)&[1.2,\,1.5)&[1.5,\,2.0)&[2.0,\,$\qsqmax$) & & \\
 Raw yield  & 251(17)  & 232(16)  & 204(15)  & 194(15)  & 161(13)  & 173(14)  & 159(13)    & &   \\
 Corrected yield  & 858(60)   & 795(59)   & 636(51)   & 612(50)   & 495(45)   & 532(46)   & 505(45)     & &   \\
 Partial rate (ns$^{-1}$) & 1.39(10)   & 1.22(9)   & 1.02(8)   & 0.98(8)   & 0.79(7)   & 0.84(7)   & 0.80(7)     & &   \\ \hline
 & \multicolumn{9}{c}{$\mathbf{\chargedkenu}$}  \rule[-1mm]{-1mm}{4.3mm} \\ $q^2$(GeV$^2/c^4$)      & [0,\,0.2) & [0.2,\,0.4) & [0.4,\,0.6) & [0.6,\,0.8) & [0.8,\,1.0) & [1.0,\,1.2) & [1.2,\,1.4) & [1.4,\,1.6) & [1.6,\,$\qsqmax$) \\
 Raw yield  & 2751(54)  & 2541(51)  & 2325(49)  & 2034(46)  & 1662(42)  & 1243(36)  & 904(31)  & 496(23)  & 167(13)      \\
 Corrected yield  & 11290(233)  & 10026(222)  & 8629(203)  & 7432(186)  & 5867(163)  & 4485(143)  & 3393(125)  & 1983(98)  & 781(68)       \\
 Partial rate (ns$^{-1}$) & 17.82(36)   & 15.83(35)   & 13.91(32)   & 11.69(29)   & 9.36(26)   & 7.08(22)   & 5.34(19)   & 3.09(15)   & 1.28(11)    \\ \hline
 & \multicolumn{9}{c}{$\mathbf{\neutralpienu}$}  \rule[-1mm]{-1mm}{4.3mm} \\ $q^2$ (GeV$^2/c^4$) & [0,\,0.3) &[0.3,\,0.6)&[0.6,\,0.9)&[0.9,\,1.2)&[1.2,\,1.5)&[1.5,\,2.0)&[2.0,\,$\qsqmax$) & & \\
 Raw yield  & 148(13)  & 141(13)  & 124(12)  & 122(12)  & 100(11)  & 107(12)  & 96(13)    & &   \\
 Corrected yield  & 799(80)  & 748(82)  & 665(80)  & 640(77)  & 570(80)  & 625(82)  & 460(86)    & &   \\
 Partial rate (ns$^{-1}$)  & 0.71(7)   & 0.66(7)   & 0.56(6)   & 0.57(6)   & 0.48(6)   & 0.54(7)   & 0.37(7)     & &   \\ \hline
 & \multicolumn{9}{c}{$\mathbf{\neutralkenu}$}  \rule[-1mm]{-1mm}{4.3mm} \\ $q^2$(GeV$^2/c^4$)      & [0,\,0.2) & [0.2,\,0.4) & [0.4,\,0.6) & [0.6,\,0.8) & [0.8,\,1.0) & [1.0,\,1.2) & [1.2,\,1.4) & [1.4,\,1.6) & [1.6,\,$\qsqmax$) \\
 Raw yield  & 1704(42)  & 1511(40)  & 1389(38)  & 1229(36)  & 912(31)  & 809(29)  & 514(24)  & 275(17)  & 124(12)    \\
 Corrected yield  & 10090(282)  & 8732(271)  & 7934(256)  & 6951(240)  & 5101(207)  & 4511(190)  & 2812(152)  & 1412(106)  & 625(73)     \\
 Partial rate (ns$^{-1}$) & 17.79(47)   & 15.62(45)   & 14.02(43)   & 12.28(40)   & 8.92(34)   & 8.17(32)   & 4.96(25)   & 2.67(18)   & 1.19(13)    \\
 \hline\hline
 \end{tabular}
\end{table*}
\endgroup

\subsection{Partial Rate Results}
\begin{figure*}[bptb]
  \includegraphics*[width=6in]{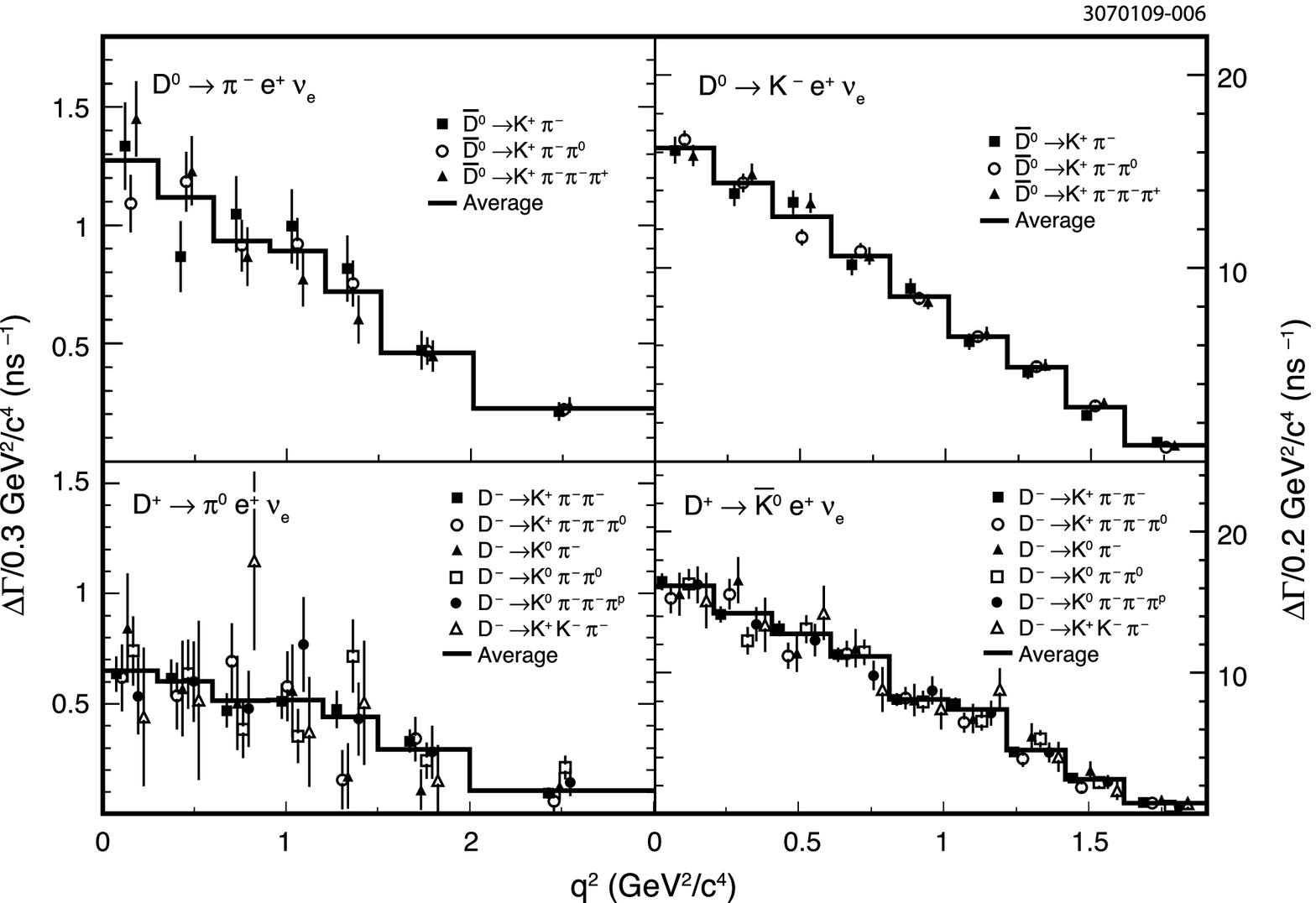}
  \caption{Partial rates for each semileptonic mode.  The points show measurements in each tag mode; the histograms show the partial rates averaged over all tag modes.}
  \label{fig:rate_4modes}
\end{figure*}

Using the tag yields, tag efficiencies, signal yields, and signal
efficiency matrices, we solve Eq.~(\ref{eq:dgdef}) for the partial rates in each $q^2$ bin and tag mode, $\Delta\Gamma_i^\alpha$.  The procedure detailed in~\cite{Lefebvre:1999yu} is used to calculate uncertainties and correlations in the inverted efficiency matrices.  We then average the resulting $\Delta\Gamma^{\alpha}_{i}$ over tag modes, obtaining $\Delta\Gamma_i$.  Statistical covariance matrices detailing the uncertainties on the $\Delta\Gamma_i$ are also calculated and are available in the Appendix.  Within each semileptonic mode, there are small correlations across $q^2$ bins that arise from the smearing in $q^2$.
  Both the individual and tag-averaged partial rates are shown in
Fig.~\ref{fig:rate_4modes}.
The tag-averaged partial rates are also given in Table~\ref{tab:sigyields}.

\subsection{Tests of Partial Rate Results}
Calculating the partial rates separately for each
tag mode allows for a test of the consistency of results
across different tag modes.  To quantify the tag mode agreement, we calculate
a $\chi^2$ for each semileptonic mode:
\begin{equation}
  \chi^2(\Gamma) = \sum_\alpha\sum_i{\frac{\left(\Delta\Gamma_i^\alpha-\Delta\Gamma_i\right)^2}{\left(\sigma^\alpha_i\right)^2}},
\end{equation}
where $\sigma^\alpha_i$ is the statistical uncertainty on
$\Delta\Gamma^\alpha_i$.
This quantity is expected to have a $\chi^2$ distribution, with mean $n_{\rm dof}$ and variance $2n_{\rm dof}$,
where the number of degrees of freedom is given by $n_{\rm dof} = n_{q^2\;{\mathrm {bins}}}\times\left(n_{\mathrm {tag\; modes}}-1\right)$.
Table~\ref{tab:variances} gives the measured $\chi^2$, the
number of degrees of freedom, and the $\chi^2$ probability for each mode.  These values show that the semileptonic rates agree well across tag modes.

\begin{table}[bptb]
\caption{$\chi^2$ of partial rates across tag modes, with number of degrees of freedom and $\chi^2$ probability.}
  \begin{center}
    \begin{tabular}{cccc}
      \hline\hline
      Semileptonic mode & $\chi^2$ & $n_{\rm dof}$ & $P(\chi^2)$ \rule[-1mm]{-1mm}{4.3mm}\\ \hline
      $\chargedpienu$ & $12$ & $14$ & 61\% \rule[-1mm]{-1mm}{4.3mm}\\
      $\chargedkenu$ & $21$  & $18$ & 28\%\\
      $\neutralpienu$ & $36$ & $35$ & 42\%\\
      $\neutralkenu$ & $37$  & $45$ & 80\%\\
      \hline\hline
    \end{tabular}
  \end{center}
  \label{tab:variances}
\end{table}

\begin{figure}[bptb]
  \includegraphics*[width=3in]{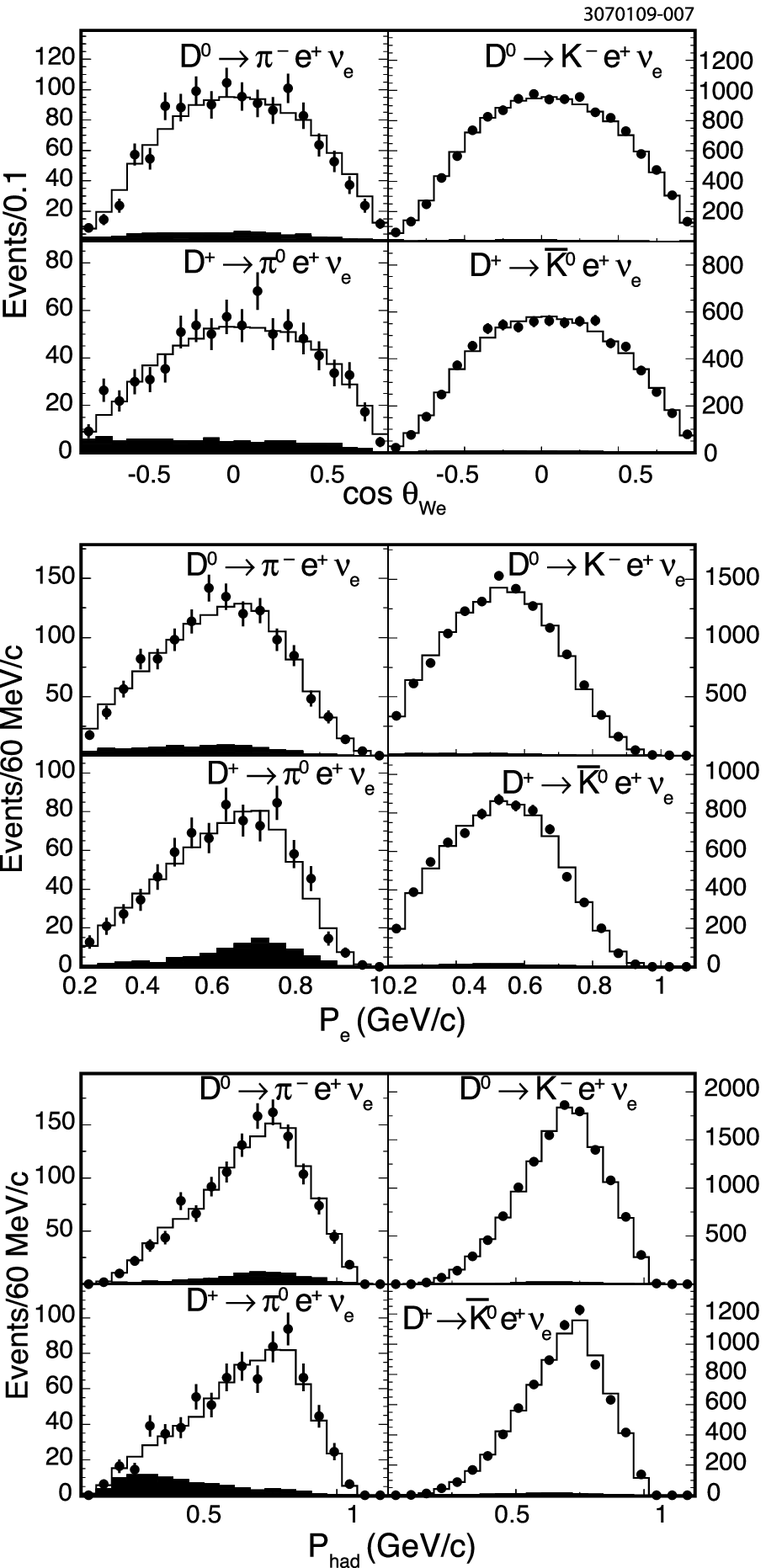}
  \caption{Distributions of $\cos\theta_{We}$ (upper), the cosine of the angle between the virtual $W$ and the positron, positron momentum (middle) and hadron momentum (lower) in events satisfying $-60$ MeV $<U< 60$ MeV.  Signal and background shapes are taken from MC simulations and scaled using the parameters of the signal yield fits.}
  \label{fig:altdists}
\end{figure}

As a test of the signal yield fits, we compare the observed and predicted distributions in three variables: the lepton momentum,  the hadron momentum, and the angle of the virtual $W$ in the $D$ rest frame relative to the positron in the $W$ rest frame.  Fig.~\ref{fig:altdists} shows these distributions for data and MC candidates with $\left|U\right|<60$ MeV, with signal and background MC distributions normalized using the $U$ fits.  All of the distributions show good agreement between data and MC simulations.

\begin{figure}[bptb]
  \includegraphics*[width=3in]{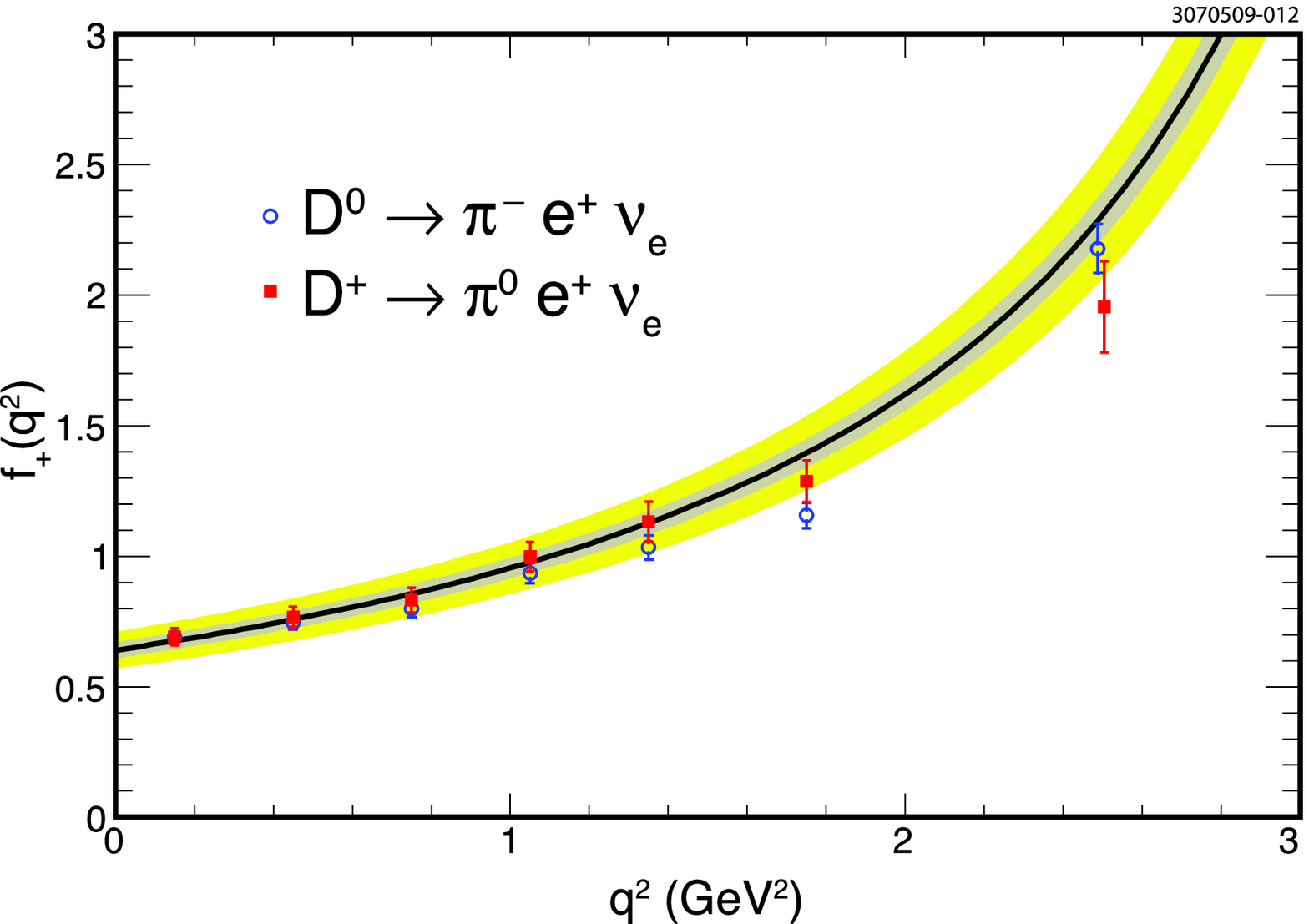}
  \includegraphics*[width=3in]{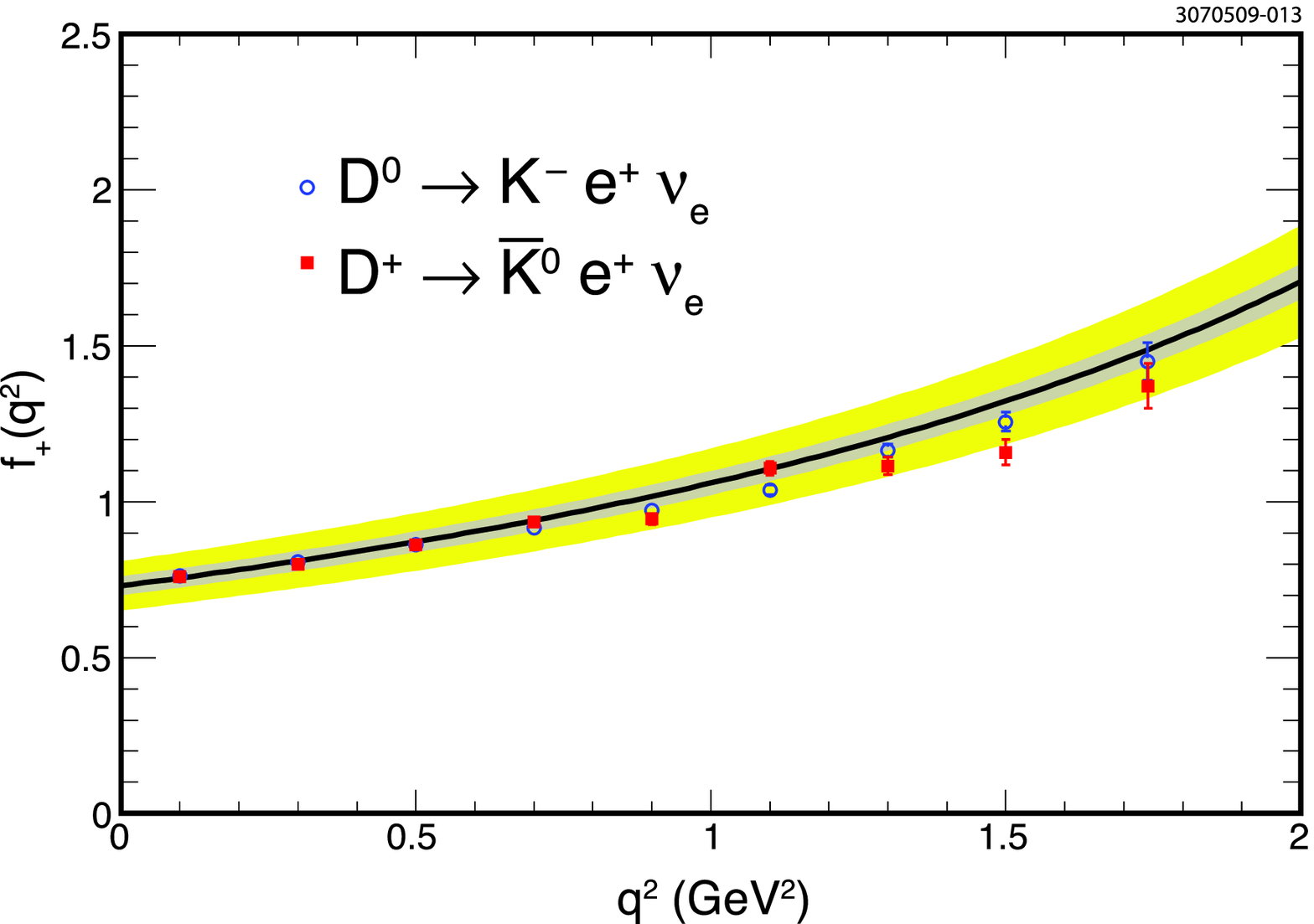}
  \caption{$f_+(q^2)$ comparison between isospin conjugate modes
  and with LQCD calculations~\cite{Aubin:2004ej}.
    The solid lines represent LQCD fits to the modified pole model~\cite{Becirevic:1999kt}.
    The inner bands show LQCD statistical uncertainties, and the outer bands the sum in quadrature of LQCD statistical and systematic uncertainties.}
  \label{fig:isospin}
\end{figure}

We also check consistency between isospin conjugate pairs.  Isospin symmetry implies that total rates for $\chargedkenu$ and $\neutralkenu$ are approximately equal, while the total rate for $\chargedpienu$ is approximately twice that of $\neutralpienu$.  After correcting for phase space differences, our partial rates summed over all $q^2$ bins agree with these expectations within 1.4 standard deviations.  Because there are small differences in phase space, it is convienient to compare not rates, but form factors, as shown in Fig.~\ref{fig:isospin}.
We obtain the $f_+(q^2)$ at the center of $q^2$ bin $i$ using
\begin{equation}
f_+ (q^2_{i})
= \frac{1}{|\vcq|}\cdot \sqrt {  \frac{\Delta \Gamma_i}{\Delta q^2_i} \frac {
 24 \pi^3  } {G^2_F   p^{3}_{i} } },
\end{equation}
where $\Delta q^2_i$ is the size of $q^2$ bin $i$,
$|\vcd| = 0.2256\pm0.0010$ and $|\vcs| = 0.97334\pm0.00023$ are from Particle Data Group fits assuming CKM unitarity~\cite{Amsler:2008zz},
and the effective $p^3$ in $q^2$ bin $i$ is given by
\begin{equation}
p^{3}_{i} =
\frac{\displaystyle \int_{i} p^{3} |f_+ (q^2)|^2 d q^2 }{|f_+(q^2_{i})|^2 \Delta q^2_i},
\end{equation}
where $f_+(q^2)$ and $f_+(q^2_i)$ are calculated using the three parameter series parameterization
with parameters measured in the data (see Sec.~\ref{sec:ff_results}).

Our measured form factors in each $q^2$ bin are seen to be in good agreement with the LQCD calculations~\cite{Aubin:2004ej},
but with significantly smaller uncertainties,
as shown in Fig.~\ref{fig:isospin}.

The procedure for measuring partial rates is tested using the generic MC sample,
from which events are drawn randomly to form mock data samples, each equivalent in size
to the data sample. In each case, the measured partial rates are
consistent with the input rates and the distributions of the deviations
are consistent with Gaussian statistics.

\section{Systematic Uncertainties in Partial Rates}
\label{sec:systematics}
Our determinations of the $\Delta \Gamma_i$ are subject to a variety of systematic uncertainties.   Tables~\ref{tab:systable_rates1} and \ref{tab:systable_rates2} list each source of systematic uncertainty and its contribution to the total uncertainty in each of the partial rates.  Because we are interested in measuring form factor shapes that vary with $q^2$, it is important that we not only understand the uncertainties in the individual partial rates but also their correlations across $q^2$.  For each semileptonic mode and each significant source of systematic uncertainty, we construct an $m\times m$ (where $m$ is the number of $q^2$ bins studied for the mode in question) covariance matrix that encapsulates both of these pieces of information.  We now describe how each of the covariance matrices is estimated.

\begingroup
\squeezetable
\begin{table*}
\caption{Summary of partial rate ($\Delta\Gamma_i$) uncertainties in percent for $\chargedpienu$ and $\chargedkenu$.  The sign gives the direction of change relative to the change in the first $q^2$ bin.}
\begin{tabular}{lccccccccc}
\hline\hline
& $\sigma(\Delta\Gamma_1)$  & $\sigma(\Delta\Gamma_2)$   & $\sigma(\Delta\Gamma_3)$    & $\sigma(\Delta\Gamma_4)$   & $\sigma(\Delta\Gamma_5)$   & $\sigma(\Delta\Gamma_6)$    & $\sigma(\Delta\Gamma_7)$   & $\sigma(\Delta\Gamma_8)$   & $\sigma(\Delta\Gamma_9)$   \rule[-1mm]{-1mm}{4.3mm} \\ \hline
$\chargedpienu$ &  \multicolumn{9}{c}{} \rule[-1mm]{-1mm}{4.3mm} \\
Tag line shape  & 0.40  & 0.40  & 0.40  & 0.40  & 0.40  & 0.40  & 0.40 & &  \\
Tag fakes  & 0.40  & 0.40  & 0.40  & 0.40  & 0.40  & 0.40  & 0.40 & &  \\
Tracking efficiency  & 0.48  & 0.48  & 0.48  & 0.48  & 0.48  & 0.49  & 0.51 & &  \\
$\pi^\pm$ ID  & 0.21  & 0.11  & 0.05  & 0.03  & 0.02  & 0.02  & 0.04 & &  \\
$e^\pm$ ID  & 0.37  & 0.38  & 0.38  & 0.39  & 0.33  & 0.18  & -0.14 & &  \\
FSR  & 0.18  & 0.11  & 0.09  & -0.02  & -0.10  & -0.20  & -0.24 & &  \\
Signal shape  & 0.56  & 0.46  & 0.58  & 0.49  & 0.50  & 0.56  & 0.49 & &  \\
Backgrounds  & 0.39  & 0.43  & 0.60  & 0.61  & 0.58  & 0.52  & 0.76 & &  \\
MC form factor  & 0.06  & -0.05  & -0.05  & -0.05  & -0.07  & -0.11  & -0.04 & &  \\
$q^{2}$ smearing  & 0.84  & -0.11  & -0.26  & -0.16  & 0.30  & -0.60  & -0.28 & &  \\
D Lifetime  & 0.37  & 0.37  & 0.37  & 0.37  & 0.37  & 0.37  & 0.37 & &  \\
All systematic  & 1.44  & 1.13  & 1.27  & 1.22  & 1.22  & 1.31  & 1.30 & &  \\
Statistical
  & 6.84  & 7.29  & 7.90  & 8.06  & 8.87  & 8.42  & 8.63 & & \\ \hline
$\chargedkenu$ &  \multicolumn{9}{c}{} \rule[-1mm]{-1mm}{4.3mm} \\
Tag line shape  & 0.40  & 0.40  & 0.40  & 0.40  & 0.40  & 0.40  & 0.40  & 0.40  & 0.40 \\
Tag fakes  & 0.40  & 0.40  & 0.40  & 0.40  & 0.40  & 0.40  & 0.40  & 0.40  & 0.40 \\
Tracking efficiency  & 0.69  & 0.72  & 0.75  & 0.79  & 0.84  & 0.92  & 1.04  & 1.26  & 1.22 \\
$K^\pm$ ID  & 0.19  & 0.13  & 0.11  & 0.10  & 0.09  & 0.10  & 0.11  & 0.11  & 0.21 \\
$e^\pm$ ID  & 0.41  & 0.42  & 0.43  & 0.45  & 0.47  & 0.48  & 0.44  & 0.33  & 0.21 \\
FSR  & 0.12  & 0.08  & 0.07  & 0.01  & -0.10  & -0.15  & -0.23  & -0.28  & -0.32 \\
Signal shape  & 0.16  & 0.12  & 0.12  & 0.14  & 0.12  & 0.11  & 0.09  & 0.14  & 0.21 \\
Backgrounds  & 0.14  & 0.04  & 0.12  & 0.09  & 0.08  & 0.08  & 0.04  & 0.10  & 0.33 \\
MC form factor  & 0.02  & -0.02  & -0.02  & -0.01  & -0.01  & -0.01  & 0.00  & 0.02  & -0.08 \\
$q^{2}$ smearing  & 0.62  & -0.11  & 0.07  & -0.12  & -0.06  & -0.51  & 0.08  & -0.62  & -2.05 \\
D Lifetime  & 0.37  & 0.37  & 0.37  & 0.37  & 0.37  & 0.37  & 0.37  & 0.37  & 0.37 \\
All systematic  & 1.26  & 1.10  & 1.12  & 1.15  & 1.20  & 1.36  & 1.35  & 1.63  & 2.55 \\
Statistical
  & 2.03  & 2.19  & 2.31  & 2.47  & 2.73  & 3.14  & 3.63  & 4.90  & 8.43\\ \hline
\hline
\end{tabular}
\label{tab:systable_rates1}
\end{table*}
\endgroup

\begingroup
\squeezetable
\begin{table*}
\caption{Summary of partial rate ($\Delta\Gamma_i$) uncertainties in percent for $\neutralpienu$ and $\neutralkenu$.  The sign gives the direction of change relative to the change in the first $q^2$ bin.}
\begin{tabular}{lccccccccc}
\hline\hline
& $\sigma(\Delta\Gamma_1)$  & $\sigma(\Delta\Gamma_2)$   & $\sigma(\Delta\Gamma_3)$    & $\sigma(\Delta\Gamma_4)$   & $\sigma(\Delta\Gamma_5)$   & $\sigma(\Delta\Gamma_6)$    & $\sigma(\Delta\Gamma_7)$   & $\sigma(\Delta\Gamma_8)$   & $\sigma(\Delta\Gamma_9)$   \rule[-1mm]{-1mm}{4.3mm} \\ \hline
$\neutralpienu$ &  \multicolumn{9}{c}{} \rule[-1mm]{-1mm}{4.3mm} \\
Tag line shape  & 0.40  & 0.40  & 0.40  & 0.40  & 0.40  & 0.40  & 0.40 & &  \\
Tag fakes  & 0.70  & 0.70  & 0.70  & 0.70  & 0.70  & 0.70  & 0.70 & &  \\
Tracking efficiency  & 0.25  & 0.25  & 0.25  & 0.24  & 0.24  & 0.24  & 0.23 & &  \\
$\pi^0$ ID  & 1.06  & 0.98  & 1.04  & 1.22  & 1.83  & 2.14  & 1.96 & &  \\
$e^\pm$ ID  & 0.32  & 0.32  & 0.34  & 0.32  & 0.27  & 0.13  & -0.22 & &  \\
FSR  & 0.14  & 0.20  & 0.08  & -0.05  & -0.14  & -0.22  & -0.21 & &  \\
Signal shape  & 1.72  & 0.93  & 1.91  & -1.24  & 3.51  & 2.43  & 3.26 & &  \\
Backgrounds  & 0.92  & 0.82  & -1.01  & 0.72  & 0.74  & 1.38  & -6.04 & &  \\
MC form factor  & 0.15  & -0.03  & -0.07  & -0.06  & -0.10  & -0.15  & 0.57 & &  \\
$q^{2}$ smearing  & 1.69  & 0.28  & -1.74  & 1.45  & -0.17  & -1.22  & -1.41 & &  \\
D Lifetime  & 0.67  & 0.67  & 0.67  & 0.67  & 0.67  & 0.67  & 0.67 & &  \\
All systematic  & 3.01  & 1.97  & 3.17  & 2.63  & 4.18  & 3.89  & -7.38 & &  \\
Statistical
  & 9.25  & 10.23  & 11.24  & 11.28  & 13.44  & 12.38  & 17.98 & & \\ \hline
$\neutralkenu$ &  \multicolumn{9}{c}{} \rule[-1mm]{-1mm}{4.3mm} \\
Tag line shape  & 0.40  & 0.40  & 0.40  & 0.40  & 0.40  & 0.40  & 0.40  & 0.40  & 0.40 \\
Tag fakes  & 0.70  & 0.70  & 0.70  & 0.70  & 0.70  & 0.70  & 0.70  & 0.70  & 0.70 \\
Tracking efficiency  & 0.76  & 0.77  & 0.78  & 0.79  & 0.81  & 0.83  & 0.87  & 0.91  & 0.96 \\
$K^0$ ID  & 2.00  & 1.96  & 1.90  & 1.83  & 1.71  & 1.51  & 1.25  & 1.35  & 1.89 \\
$e^\pm$ ID  & 0.42  & 0.43  & 0.43  & 0.45  & 0.48  & 0.48  & 0.44  & 0.33  & 0.20 \\
FSR  & 0.17  & 0.13  & 0.08  & 0.01  & -0.11  & -0.16  & -0.23  & -0.24  & -0.28 \\
Signal shape  & 0.20  & 0.22  & 0.17  & 0.20  & 0.23  & 0.26  & 0.38  & 0.26  & 0.47 \\
Backgrounds  & 0.13  & 0.13  & 0.11  & 0.11  & 0.14  & 0.15  & 0.27  & 0.23  & 1.46 \\
MC form factor  & 0.03  & -0.02  & -0.02  & -0.02  & -0.02  & -0.01  & 0.01  & 0.02  & 0.08 \\
$q^{2}$ smearing  & 0.63  & -0.24  & -0.02  & 0.29  & -1.06  & 0.75  & -0.67  & -0.78  & -1.11 \\
D Lifetime  & 0.67  & 0.67  & 0.67  & 0.67  & 0.67  & 0.67  & 0.67  & 0.67  & 0.67 \\
All systematic  & 2.52  & 2.42  & 2.36  & 2.33  & 2.47  & 2.23  & 2.08  & 2.16  & 3.03 \\
Statistical
  & 2.63  & 2.90  & 3.04  & 3.23  & 3.82  & 3.98  & 5.04  & 6.88  & 10.63\\ \hline
\hline
\end{tabular}
\label{tab:systable_rates2}
\end{table*}
\endgroup

Tag reconstruction biases enter both the numerator and denominator of our partial rate formulation in Eq.~(\ref{eq:dgdef}), and therefore largely cancel.  However, there are two sources of systematic uncertainty related to tag yields.  One source originates in the line shapes used to extract tag yields in data; we estimate this by using alternate line shapes and find an uncertainty of 0.4\% for partial rates in both $D^0$ and $D^-$ modes.  The selection of one tag per mode also introduces a systematic uncertainty, primarily due to possible mismodeling of MC $\pi^0$ fake rates.  Based on estimates of tag-fake rates in data and MC samples, we assign a systematic uncertainty of 0.4\% to the partial rates in $D^0$ modes and 0.7\% to those in $D^-$ modes, where a greater fraction of tags contain $\pi^0$'s.  As the uncertainties associated with tag yields are independent of the kinematics of the semileptonic decay, they are fully correlated across $q^2$ bins.

Systematic uncertainties associated with semileptonic track, $\pi^-$, $K^-$, $\pi^0$, and $K^0_S$ reconstruction are all studied in a similar manner: we choose fully hadronic events containing a particle of type $X$, where $X=\pi^-$, $K^-$, $\pi^0$, or $K^0_S$, and reconstruct all particles in the event except for $X$.  We then form missing mass squared distributions, which peak at $M_X^2$ for correctly reconstructed events.  We then tally the fraction of events with the appropriate $M_X^2$ in which $X$ was successfully reconstructed, after correcting for backgrounds.  By doing this in bins of missing momentum, we compare the data and MC efficiencies as a function of particle momentum.

In the case of $\pi$ and $K$ track reconstruction and $K^0_S$ finding, no evidence of bias in the efficiencies is found.  Biases of less than 1\% are observed in $\pi^-$ and $K^-$ identification efficiencies.  We also find $\pi^0$ reconstruction efficiencies to be approximately 6\% lower in data than in MC simulations, the bias being roughly constant across $\pi^0$ momentum.  About half of this discrepancy has been traced to incorrect modeling of the lateral spread of photon showers in the calorimeter and the energy resolution; the other half is of unknown origin.  We reweight the MC distributions to correct for all reconstruction biases.

Systematic uncertainties in the particle reconstruction efficiencies are often correlated across momentum bins.  When this is the case, we construct
a covariance matrix binned in particle momentum by noting that efficiencies binned in particle momentum ($\epsilon_{X}^p$) are related to efficiencies binned in semileptonic $q^2$ ($\epsilon_{X}^{q^2}$) via
\begin{equation}
\epsilon_{X}^{q^2} = {\mathbf{A}}\epsilon_{X}^{p},
\end{equation}
where $\mathbf{A}$ is a matrix giving the fraction of type $X$ particles that are part of a semileptonic decay in a given $q^2$ bin that are also in a given momentum bin; this matrix is estimated using signal MC simulations.  The fractional covariance matrix binned in $q^2$, ${\mathbf{M}}^{q^2}$, is then given by
\begin{equation}
\mathbf{M}^{q^2}=\mathbf{A}\mathbf{M}^p\mathbf{A}^{\rm{T}},
\end{equation}
where $\mathbf{M}^{p}$ is the momentum-binned fractional covariance matrix.  This equation is used to obtain the $q^2$ binned systematic covariance matrices associated with track, $\pi^-$, $K^-$, $\pi^0$, and $K_S^0$ reconstruction.

The covariance matrices for all remaining systematic uncertainties -- those associated with positron identification, FSR, background and signal shapes used to obtain signal yields, form factor parameterization in MC simulations, and smearing in $q^2$ -- are estimated by the following procedure: for each source of systematic uncertainty, we vary the analysis in a manner that approximates the uncertainty on the effect in question and remeasure the partial rates $\Delta\Gamma_i$.  The covariance matrix elements $M_{ij}$ for this source can then be estimated via
\begin{equation}
  \mathbf{M}_{ij} = \delta\left(\Delta\Gamma_i\right)\delta\left(\Delta\Gamma_j\right),
\end{equation}
where $\delta\left(\Delta\Gamma_i\right)$ denotes the difference between the partial rate in $q^2$ bin $i$ measured using the varied analysis and the rate using the standard analysis technique.  In most cases, we make several variations to the analysis and sum the resulting covariance matrices.  Where it is possible to vary some parameter by positive and negative values, we average the results of the positive and negative variations.

Positron identification efficiencies as a function of positron momentum are measured in MC simulations and in data using radiative Bhabha ($ee\gamma$) and two-photon ($eeee$) events.  Since the positrons in these events are relatively isolated, we embed these positrons into hadronic events, and determine the decrease  in efficiency.  Biases of around 1.5\% are observed, and the MC signal efficiency matrices and $U$ distributions are corrected for these biases.  To estimate the systematic uncertainty due to positron identification, we shift the corrections by the uncertainties on their measurement and remeasure the partial rates using efficiencies and $U$ distributions obtained with the shifted corrections.

FSR affects the partial rate measurements primarily by causing mismeasurements of positron momentum.  FSR in the MC simulations is modeled using {\sc PHOTOS} version 2.15, which models FSR significantly better than earlier versions.  To estimate the systematic uncertainty due to FSR, we reweight the efficiency matrices and $U$ distributions so that the energy and angular distribution of photons reconstructed in the neighborhood of the positrons match those measured in data, and remeasure the partial rates.  We have also studied systematic uncertainties associated with ISR, which are found to be negligible.

The signal shapes used to model semileptonic candidates in the signal yield fits are taken from signal MC distributions convolved with a double Gaussian.  The systematic uncertainty associated with this procedure is estimated by varying the widths of the Gaussians and the normalization of the wider one within their uncertainties and remeasuring the signal yields.  In the $\neutralpienu$ mode, there is also evidence of a possible shift between data and signal MC $U$ distributions.  In this mode only, we apply a systematic uncertainty equal to the change in rates when a shift is applied.

The background lineshapes in the signal yield fits introduce systematic uncertainties in three ways.  The first arises from our choice to fix the normalization of several background shapes, namely the small non-$D\bar{D}$ background in all modes, the $\rhoenu$ and $\chargedkenu$ backgrounds to $\chargedpienu$ and the $\neutralkenu$ background to $\neutralpienu$.  The systematic uncertainties associated with these backgrounds are estimated by varying the normalizations within their uncertainties.  In the case of the $\chargedkenu$ and $\neutralkenu$ backgrounds, where the normalizations are those that minimize the fit likelihoods summed over all $q^2$ and tag modes, we vary the normalization to values that increase the likelihood by unity.  Secondly, the choice to combine many background modes into one shape using fixed relative normalizations may result in incorrect background shapes.  We estimate this systematic uncertainty by varying the normalization of several of the largest components of the combined shapes based on branching fraction uncertainties.  Finally, incorrect MC fake rates may also lead to inaccurate background shapes.  Our technique is most sensitive to positron and $\pi^0$ fake rates.  Using estimates of hadron-to-positron fake rates studied in $D^+\rightarrow K^-\pi^+\pi^+$ and $K^0_S\rightarrow\pi^+\pi^-$ and $\pi^0$ fake rates studied in $\kpipiz$, we estimate this systematic uncertainty by increasing the fake rates in MC simulations to match those found in data.

The use of efficiency and smearing matrices binned in $q^2$ reduces the dependence of our results on the $f_+(q^2)$ used to generate signal events in the MC simulations.  However, we are still sensitive to non-linear effects within $q^2$ bins.  To estimate the systematic uncertainty associated with this effect, we reweight signal MC events so that the $q^2$ spectra follow alternate form factor parameterizations.  These variations sample the one standard deviation ellipsoid of the form factor measurements reported in this article.

We account for possible mismodeling of the resolution in $q^2$.
Using $q^2=m_D^2+m_{h}^2-2m_D E_{h}$, where $E_h$ is the hadron energy, it follows that $\delta q^2=-2m_D \delta E_{h}$.
Assuming the $U$ resolution is dominated by the hadron energy resolution,
we estimate that the $q^2$ resolution may be 0.05 GeV$^2/c^4$ larger in data than in MC simulations in $\neutralpienu$ and 0.02 GeV$^2/c^4$ larger in the other modes.  Smearing the MC $q^2$ distributions by these amounts leads to the changes in the partial rates shown in Tables~\ref{tab:systable_rates1} and \ref{tab:systable_rates2} .  A final systematic uncertainty on the partial rates arises from the $D^0$ and $D^\pm$ lifetimes.  Using PDG 2008 values~\cite{Amsler:2008zz}, these are 0.37\% and 0.67\%, respectively.

\section{Form Factor, Branching Fraction, and CKM Measurements}
\label{sec:ff_fits}
To extract form factor parameters, branching fractions, and the magnitudes of CKM elements $|\vcd|$ and $|\vcs|$, we fit the partial rate results using Eq.~(\ref{eq:diffrate}) and parameterizations of the form factor $f_+\left(q^2\right)$.  Several parameterizations have been suggested.  We now review these, before reporting the results of our fits.

\subsection{Form Factor Parameterizations}
\label{sec:ff_params}
While the exact form of $f_+\left(q^2\right)$ is not calculable in QCD, some information about the form factor is available.  Specifically, it is expected to be an analytic fuction everywhere in the complex $q^2$ plane outside of a cut that extends along the positive $q^2$ axis from the mass of the lowest-lying $c\bar{d}$ (for $D\rightarrow\pi$) or $c\bar{s}$ (for $D\rightarrow K$) vector meson.  This assumption leads to a dispersion relation (see for example Ref.~\cite{Becher:2005bg}):
\begin{eqnarray}
f_+\left(q^2\right) & = & \frac{f_+\left(0\right)/(1-\alpha)}{1-\frac{\qsq}{M^2_{D^*_{(s)}}}} \nonumber \\
 & + & \frac{1}{\pi}\int_{(m_D+m_P)^2}^\infty {\frac{{\rm Im}f_+\left(t\right)}{t-\qsq-i\epsilon}dt},
\label{eq:dispersion}
\end{eqnarray}
where $m_D$ and $m_P$ are the masses of the semileptonic parent and daughter mesons, respectively, $m_{D^*_{(s)}}$ is the mass of the $D^*$ for $D\rightarrow\pi$ or $D^*_s$ for $D\rightarrow K$, and $\alpha$ gives the relative contribution of this meson to $f_+\left(0\right)$.  Most of the suggested form factor parameterizations are motivated by a version of this dispersion relation where the integral has been replaced by a sum over effective poles:
\begin{eqnarray}
f_+\left(q^2\right)& = & \frac{f_+\left(0\right)/(1-\alpha)}{1-\frac{\qsq}{M^2_{D^*_{(s)}}}} \nonumber \\
 & + & \sum_{k=1}^{N}{\frac{\rho_k}{1-\frac{1}{\gamma_k}\frac{q^2}{M_{D^*_{(s)}}^2}}},
\label{eq:disp_rel_sum}
\end{eqnarray}
where the expansion parameters $\rho_k$ and $\gamma_k$ are unknown.

A parameterization known as the simple pole model assumes that the sum in Eq.~(\ref{eq:disp_rel_sum}) is dominated by a single pole~\cite{Becirevic:1999kt}:
\begin{equation}
f_+(q^2) = \frac{f_+(0)}{1-\frac{q^2}{m_{\rm pole}^2}}.
\end{equation}
where the value of $m_{\rm pole}$ is predicted to be $M_{D^*_{(s)}}$.

Another parameterization, known as the modified pole model \cite{Becirevic:1999kt}, adds a second term to the expansion given in Eq.~(\ref{eq:disp_rel_sum}), thus assuming that all higher order poles can be modeled by a single effective pole.  To reduce the number of free parameters, this model also makes several simplifications, including assumptions that $\beta$, a parameter that quantifies scaling violations, is near unity and $\delta$, which describes the hard scattering of gluons, is near zero, leading to the prediction that
\begin{equation}
1+1/\beta-\delta\equiv{\frac{m_D^2-m_P^2}{f_+(0)}\frac{df_+(q^2)}{dq^2}}\bigg|_{q^2=0}\approx 2.
\label{eq:betadelta}
\end{equation}
After making these simplifying assumptions, the two pole terms are reduced to
\begin{equation}
f_+(q^2) = \frac{f_+(0)}{(1-\frac{q^2}{m_{\rm pole}^2})(1-\alpha\frac{q^2}{m_{\rm pole}^2})},
\end{equation}
where $m_{\rm pole}$ is generally fixed to the $D^*_{(s)}$ mass and $\alpha$ is a free  parameter.

While the simple and modified pole parameterizations have been widely used, the presence of poles near the semileptonic $q^2$ regions causes the sum in Eq.~(\ref{eq:disp_rel_sum}) to have poor convergence properties, creating doubt as to whether truncating all but the first one or two terms leaves an accurate estimate of the true form factor.  A third parameterization, known as the series expansion, attempts to address the problem~\cite{Becher:2005bg,Hill:2006ub,Hill:2007xc}.  Exploiting the analytic properties of $f_+\left(q^2\right)$, a transformation of variables is made that maps the cut in the $q^2$ plane onto a unit circle $\left|z\right|<1$, where
\begin{equation}
  z(q^2,t_0) = \frac{\sqrt{t_+-\qsq}-\sqrt{t_+-t_0}}{\sqrt{t_+-q^2}+\sqrt{t_+-t_0}},
\end{equation}
$t_\pm=(m_D\pm m_P)^2$, and $t_0$ is any real number less than $t_+$.  This transformation amounts to expanding the form factor about $q^2=t_0$, with the expanded form factor given by
\begin{equation}
f_+(q^2) = \frac{1}{P(q^2)\phi(q^2,t_0)}\sum_{k=0}^\infty a_k\left(t_0\right)\left[z\left(q^2,t_0\right)\right]^k,
\label{eq:series_expansion}
\end{equation}
where $a_k$ are real coefficients, $P(q^2)=z(q^2,M_{D^*}^2)$ for kaon final states, $P(q^2)=1$ for pion final states, and $\phi(q^2,t_0)$ is any function that is analytic outside a cut in the complex $q^2$ plane that lies along the $x$-axis from $t_+$ to $\infty$.  This expansion has improved convergence properties over Eq.~(\ref{eq:disp_rel_sum}) due to the smallness of $z$; for example, taking the traditional choice of $t_0=t_+\left(1-(1-t_-/t_+)^{1/2}\right)$, which minimizes the maximum value of $z(q^2,t_0)$, the maximum value of $z$ over the semileptonic $q^2$ region is 0.17 for $D\rightarrow\pi$ and 0.051 for $D\rightarrow K$ \cite{Hill:2006ub}.  Further, taking the standard choice of $\phi$:
\begin{eqnarray}
\phi(q^2,t_0) & = & \sqrt{\frac{\pi m_c^2}{3}}\left(\frac{z\left(q^2,0\right)}{-q^2}\right)^{5/2}\left(\frac{z\left(q^2,t_0\right)}{t_0-q^2}\right)^{-1/2} \nonumber\\
          &\times & \left(\frac{z\left(q^2,t_-\right)}{t_--q^2}\right)^{-3/4}\frac{t_+-q^2}{\left(t_+-t_0\right)^{1/4}},
\end{eqnarray}
it can be shown that the sum over all $k$ of $a_k^2$ is of order unity \cite{Hill:2007xc}.

While the three parameterizations described above are the most commonly used, a fourth parameterization, known as ISGW2~\cite{Scora:1995ty}, is also occasionally used.  Based on a quark model, this parameterization hypothesizes
\begin{equation}
  f_+\left(q^2\right)=f_+\left(q^2_{\rm max}\right)\left(1+\frac{r_{\rm ISGW2}^2}{12}\left(q^2_{\rm max}-q^2\right)\right)^{-2},
\end{equation}
and predicts $r_{\rm ISGW2}=1.12$ GeV$^{-1}c^2$.

\subsection{Fitting Technique}
Taking into account correlations across $q^2$ bins, our fits minimize
\begin{equation}
\chi^2 = \sum_{i,j=1}^m (\DG_i - g(q^2)_i)C^{-1}_{ij} (\DG_j - g(q^2)_j),
\label{eq:ffchi}
\end{equation}
where $m$ is the number of $q^2$ bins for the mode in question, $C_{ij}$ is the sum of the statistical and systematic covariance matrices for the $\DG_j$, and $g(q^2)_j$ is the predicted partial rate in the $j$th bin for the hypothesized form factor and $|\vcq|$.  For each semileptonic mode, we perform fits using each of the four parameterizations described in Sec.~\ref{sec:ff_params}, and provide two versions of the series expansion parameterization -- one with only a linear term (referred to as the two-parameter series) and one with a linear and quadratic term (referred to as the three-parameter series).   In all cases, we vary $\left|\vcq\right|\fz$ and one or more shape parameters: $r_1\equiv a_1/a_0$ and $r_2\equiv a_2/a_0$ in the three-parameter series model, $r_1$ in the two-parameter series model, $\alpha$ in the modified pole model and $m_{\rm pole}$ in the simple pole model and $r_{\rm ISGW2}$ in the ISGW2 model.  The central values of these parameters are taken from the combined statistical and systematic fit.  To separate the statistical and systematic uncertainties, we redo the fits using only statistical covariance matrices, taking the systematic uncertainty to be the difference between the combined and statistical-only fits in quadrature.

If isospin is an exact symmetry, the form factors for $\chargedpienu$ and $\neutralpienu$ are expected to be identical,
as are those for $\chargedkenu$ and $\neutralkenu$.  For this reason, we also perform combined fits to these isospin conjugate pairs.  To accomplish this, we again minimize the $\chi^2$ given in Eq.~(\ref{eq:ffchi}), now modified so that the $\DG_i$ for the isospin conjugate pairs are combined into one vector of length $2m$ and $C_{ij}$ becomes a $2m\times2m$ covariance matrix for the combined $\DG_i$.   These covariance matrices, the diagonal blocks of which form the covariance matrices used in the fits to individual semileptonic modes, are reported in the Appendix.

The fitting technique is tested by reweighting portions of signal MC samples to alternate form factor parameterizations and treating these as mock data samples.  We find that the input parameters are reproduced with no evidence of bias when the input form factor parameters are the same as those used to obtain efficiency matrices.  When the input parameters differ from the efficiency matrix paramaters, small biases are observed, and we use these to estimate the systematic uncertainties associated with the MC form factor parameterization, as described in Sec.~\ref{sec:systematics}.

\subsection{Form Factor Results}

\label{sec:ff_results}

\begingroup
\squeezetable
\begin{table*}
\caption{Results of individual form factor fits; statistical and systematic uncertainties on the least significant digits are shown in parentheses.  For the series parameterization, we provide results of $f_+(0)\vcd$, $r_1=a_1/a_0$ and $r_2=a_2/a_0$, as well as the expansion parameters $a_0$, $a_1$ and $a_2$ themselves.  The columns labeled $\rho_{ij}$($\rho$) give the correlation coefficients of the previous three (two) parameters.}
\begin{tabular}{lccccc}
\hline\hline
3 par. series &  $f_+(0)\left|\vcq\right|$  & $r_1$ & $r_2$ & $\rho_{01}$, $\rho_{02}$, $\rho_{12}$ & $\chi^2/\rm{d.o.f.}$ \rule[-1mm]{-1mm}{4.3mm}\\
$\chargedpienu$ & 0.152(5)(1) & -2.80(49)(4) & 6(3)(0) & -0.44 0.68 -0.94 & 4.6/4 \\
$\chargedkenu$ & 0.726(8)(4) & -2.65(34)(8) & 13(9)(1) & -0.22 0.64 -0.82 & 3.2/6 \\
$\neutralpienu$ & 0.146(7)(2) & -1.37(88)(24) & -4(5)(1) & -0.43 0.65 -0.96 & 0.9/4 \\
$\neutralkenu$ & 0.707(10)(9) & -1.66(44)(10) & -14(11)(1) & -0.11 0.54 -0.82 & 11.9/6 \\
 & $a_0$ & $a_1$ & $a_2$ & $\rho_{01}$, $\rho_{02}$, $\rho_{12}$ &  \\
$\chargedpienu$ & 0.071(2)(1) & -0.20(4)(0) & 0.5(2)(0) & -0.48 0.14 -0.91 &  \\
$\chargedkenu$ & 0.0264(2)(2) & -0.07(1)(0) & 0.3(2)(0) & -0.22 -0.22 -0.79 &  \\
$\neutralpienu$ & 0.074(3)(2) & -0.10(7)(2) & -0.3(4)(1) & -0.66 0.37 -0.93 &  \\
$\neutralkenu$ & 0.0258(2)(3) & -0.04(1)(0) & -0.4(3)(0) & -0.07 -0.26 -0.78 &  \\
\hline
2 par. series & $f_+(0)\left|\vcq\right|$ & $r_1$ &  & $\rho$ & $\chi^2/\rm{d.o.f.}$ \rule[-1mm]{-1mm}{4.3mm} \\
$\chargedpienu$ & 0.145(4)(1) & -1.86(18)(3) &  & 0.83 & 8.2/5 \\
$\chargedkenu$ & 0.717(6)(4) & -2.23(19)(8) &  & 0.68 & 5.4/7 \\
$\neutralpienu$ & 0.150(5)(2) & -1.94(25)(9) &  & 0.80 & 1.3/5 \\
$\neutralkenu$ & 0.716(7)(9) & -2.10(25)(8) &  & 0.62 & 13.4/7 \\
  & $a_0$ & $a_1$ &  & $\rho$ &  \\
$\chargedpienu$ & 0.071(2)(1) & -0.13(2)(0) &  & -0.89 &  \\
$\chargedkenu$ & 0.0265(2)(2) & -0.06(1)(0) &  & -0.66 &  \\
$\neutralpienu$ & 0.074(3)(2) & -0.14(2)(1) &  & -0.92 &  \\
$\neutralkenu$ & 0.0257(2)(3) & -0.05(1)(0) &  & -0.45 &  \\
\hline
Modified pole & $f_+(0)\left|\vcq\right|$ & $\alpha$ &  & $\rho$ & $\chi^2/\rm{d.o.f.}$ \rule[-1mm]{-1mm}{4.3mm}\\
$\chargedpienu$ & 0.145(4)(1) & 0.21(8)(1) &  & -0.82 & 8.4/5 \\
$\chargedkenu$ & 0.716(6)(4) & 0.31(4)(2) &  & -0.66 & 5.9/7 \\
$\neutralpienu$ & 0.150(5)(2) & 0.24(11)(4) &  & -0.77 & 1.3/5 \\
$\neutralkenu$ & 0.715(7)(9) & 0.28(6)(2) &  & -0.61 & 13.1/7 \\
\hline
Simple pole & $f_+(0)\left|\vcq\right|$ & $m_{\rm{pole}}$ &  & $\rho$ & $\chi^2/\rm{d.o.f.}$ \rule[-1mm]{-1mm}{4.3mm}\\
$\chargedpienu$ & 0.146(3)(1) & 1.91(3)(0) &  & 0.71 & 5.8/5 \\
$\chargedkenu$ & 0.720(5)(4) & 1.91(2)(1) &  & 0.59 & 3.6/7 \\
$\neutralpienu$ & 0.153(4)(2) & 1.92(4)(1) &  & 0.65 & 2.2/5 \\
$\neutralkenu$ & 0.720(6)(9) & 1.95(3)(1) &  & 0.55 & 14.7/7 \\
\hline
ISGW2 & $f_+(0)\left|\vcq\right|$ & $r$ &  & $\rho$ & $\chi^2/\rm{d.o.f.}$ \rule[-1mm]{-1mm}{4.3mm}\\
$\chargedpienu$ & 0.142(4)(1) & 2.00(9)(1) &  & -0.80 & 12.1/5 \\
$\chargedkenu$ & 0.714(5)(4) & 1.60(3)(1) &  & -0.64 & 7.5/7 \\
$\neutralpienu$ & 0.148(5)(2) & 2.02(12)(5) &  & -0.74 & 0.9/5 \\
$\neutralkenu$ & 0.713(7)(9) & 1.58(4)(1) &  & -0.59 & 12.5/7 \\
\hline\hline
\end{tabular}
\label{tab:results2}
\end{table*}
\endgroup

\begingroup
\squeezetable
\begin{table*}
\caption{Results of isospin-combined form factor fits; statistical and systematic uncertainties on the least significant digits are shown in parentheses.  For the series parameterization, we provide results of $f_+(0)\vcd$, $r_1=a_1/a_0$ and $r_2=a_2/a_0$, as well as the expansion parameters $a_0$, $a_1$ and $a_2$ themselves.  The columns labeled $\rho_{ij}$($\rho$) give the correlation coefficients of the previous three (two) parameters.}
\begin{tabular}{lccccc}
\hline\hline
3 par. Series &  $f_+(0)\left|\vcq\right|$  & $r_1$ & $r_2$ & $\rho_{01}$, $\rho_{02}$, $\rho_{12}$ & $\chi^2/\rm{d.o.f.}$ \rule[-1mm]{-1mm}{4.3mm}\\
$D\rightarrow\pi^-/\pi^{0}e^+\nu_e$ & 0.150(4)(1) & -2.35(43)(7) & 3(3)(0) & -0.43 0.67 -0.94 & 10.4/11 \\
$D\rightarrow K^-/\bar{K}^{0}e^+\nu_e$ & 0.719(6)(5) & -2.25(27)(7) & 3(7)(1) & -0.19 0.59 -0.81 & 19.1/15 \\
 & $a_0$ & $a_1$ & $a_2$ & $\rho_{01}$, $\rho_{02}$, $\rho_{12}$ &  \\
$D\rightarrow\pi^-/\pi^{0}e^+\nu_e$ & 0.072(2)(1) & -0.17(3)(1) & 0.3(2)(0) & -0.53 0.21 -0.92 &  \\
$D\rightarrow K^-/\bar{K}^{0}e^+\nu_e$ & 0.0263(1)(2) & -0.06(1)(0) & 0.1(2)(0) & -0.19 -0.21 -0.79 &  \\
\hline
2 par. Series & $f_+(0)\left|\vcq\right|$ & $r_1$ &  & $\rho$ & $\chi^2/\rm{d.o.f.}$ \rule[-1mm]{-1mm}{4.3mm}\\
$D\rightarrow\pi^-/\pi^{0}e^+\nu_e$ & 0.146(3)(1) & -1.87(15)(4) &  & 0.81 & 11.7/12 \\
$D\rightarrow K^-/\bar{K}^{0}e^+\nu_e$ & 0.717(4)(4) & -2.17(15)(6) &  & 0.59 & 19.2/16 \\
  & $a_0$ & $a_1$ &  & $\rho$ &  \\
$D\rightarrow\pi^-/\pi^{0}e^+\nu_e$ & 0.071(2)(1) & -0.13(1)(0) &  & -0.89 &  \\
$D\rightarrow K^-/\bar{K}^{0}e^+\nu_e$ & 0.0263(1)(2) & -0.056(4)(2) &  & -0.59 &  \\
\hline
Modified pole & $f_+(0)\left|\vcq\right|$ & $\alpha$ &  & $\rho$ & $\chi^2/\rm{d.o.f.}$ \rule[-1mm]{-1mm}{4.3mm}\\
$D\rightarrow\pi^-/\pi^{0}e^+\nu_e$ & 0.146(3)(1) & 0.21(7)(2) &  & -0.80 & 12.2/12 \\
$D\rightarrow K^-/\bar{K}^{0}e^+\nu_e$ & 0.716(4)(4) & 0.30(3)(1) &  & -0.57 & 19.4/16 \\
\hline
Simple pole & $f_+(0)\left|\vcq\right|$ & $m_{\rm{pole}}$ &  & $\rho$ & $\chi^2/\rm{d.o.f.}$ \rule[-1mm]{-1mm}{4.3mm}\\
$D\rightarrow\pi^-/\pi^{0}e^+\nu_e$ & 0.148(2)(1) & 1.91(2)(1) &  & 0.68 & 10.3/12 \\
$D\rightarrow K^-/\bar{K}^{0}e^+\nu_e$ & 0.721(4)(4) & 1.93(2)(1) &  & 0.51 & 19.2/16 \\
\hline
ISGW2 & $f_+(0)\left|\vcq\right|$ & $r$ &  & $\rho$ & $\chi^2/\rm{d.o.f.}$ \rule[-1mm]{-1mm}{4.3mm}\\
$D\rightarrow\pi^-/\pi^{0}e^+\nu_e$ & 0.144(3)(1) & 1.99(7)(2) &  & -0.78 & 15.9/12 \\
$D\rightarrow K^-/\bar{K}^{0}e^+\nu_e$ & 0.714(4)(4) & 1.59(2)(1) &  & -0.55 & 20.4/16 \\
\hline\hline
\end{tabular}
\label{tab:results2_isospin}
\end{table*}
\endgroup

\begin{figure*}[bptb]
  \includegraphics*[width=5.4in]{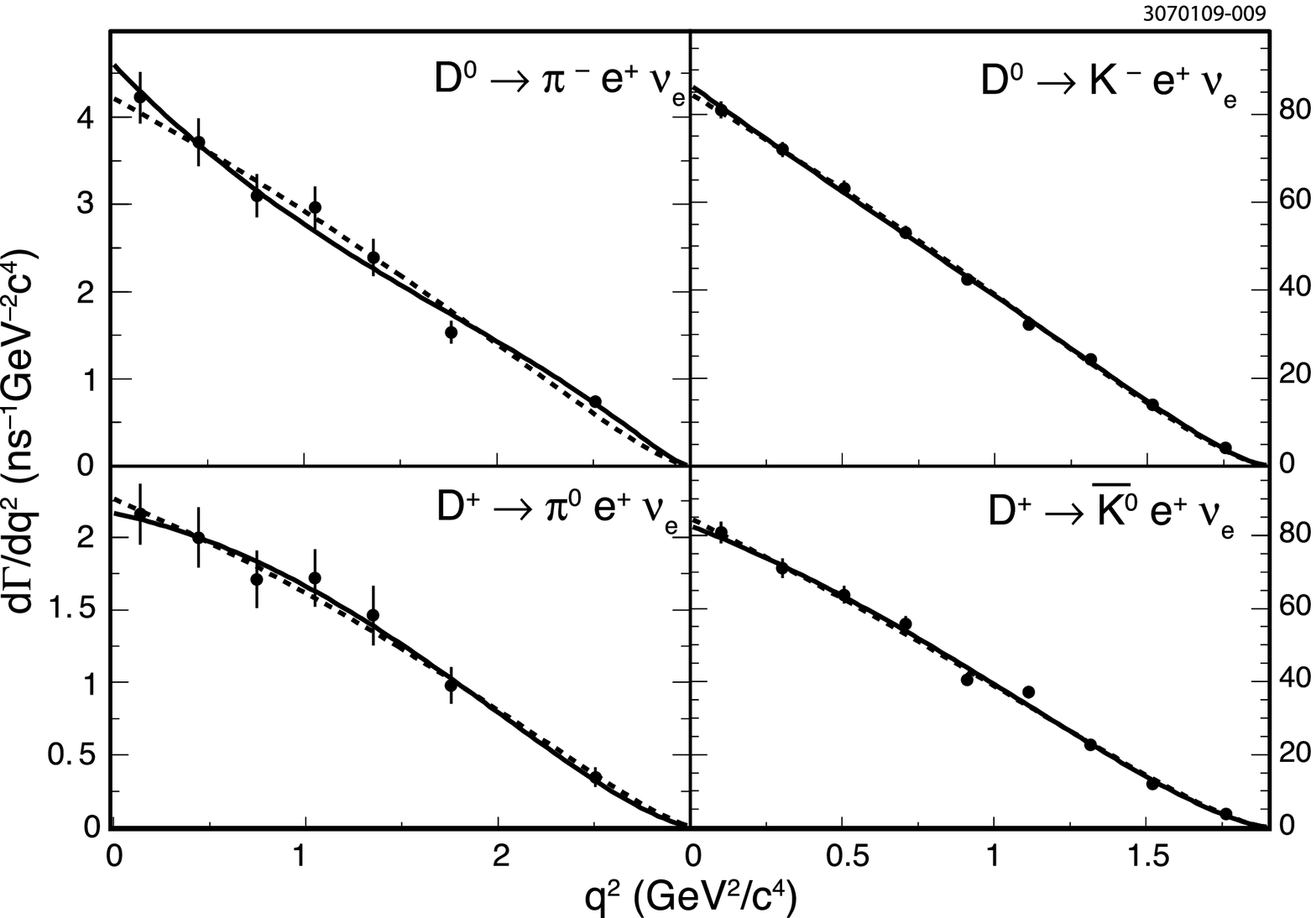}
  \caption{Individual form factor fits to data (points) using 2-parameter (dashed) and 3-parameter (solid) series expansions.}
    \label{fig:fffit}
  \includegraphics*[width=5.4in]{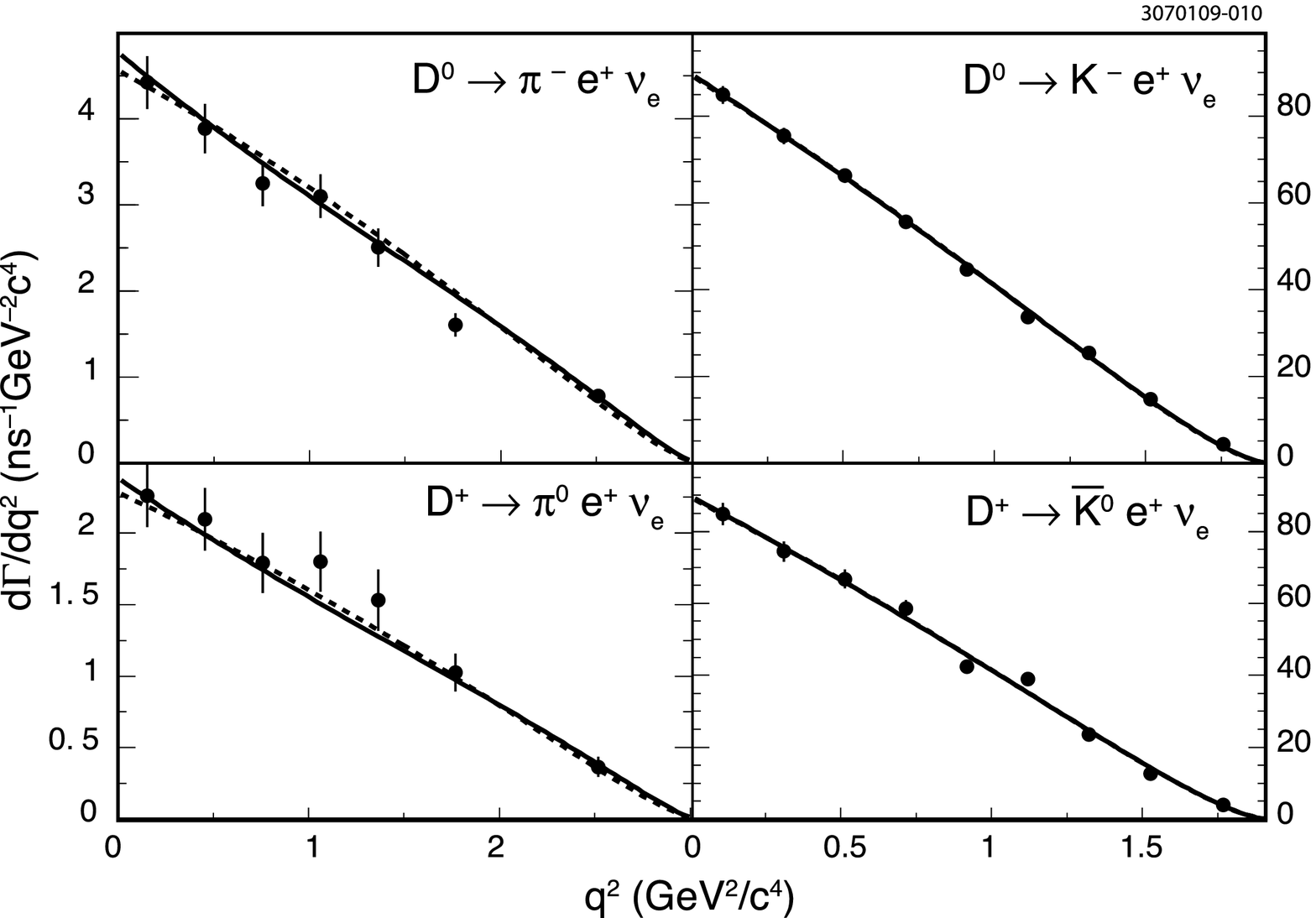}
  \caption{Isospin-combined form factor fits to data (points) using 2-parameter (dashed) and 3-parameter (solid) series expansions.}
  \label{fig:fffit_isospin}
\end{figure*}

The optimized form factor parameters, the correlations between these parameters and the minimized $\chi^2$ values from fits to each semileptonic mode using each of the parameterizations are shown in Table~\ref{tab:results2}; the corresponding values obtained from simultaneous fits to the isospin conjugate pairs are shown in Table~\ref{tab:results2_isospin}.  Plots of the three parameter series expansion fits are shown in Figs.~\ref{fig:fffit} and \ref{fig:fffit_isospin}.  To facilitate display of the fit results, we have plotted $d\Gamma/dq^2$, which is estimated by dividing the $\Delta\Gamma_i$ by the width of $q^2$ bin $i$.

\begin{figure*}[htbp]
\includegraphics*[width=6in]{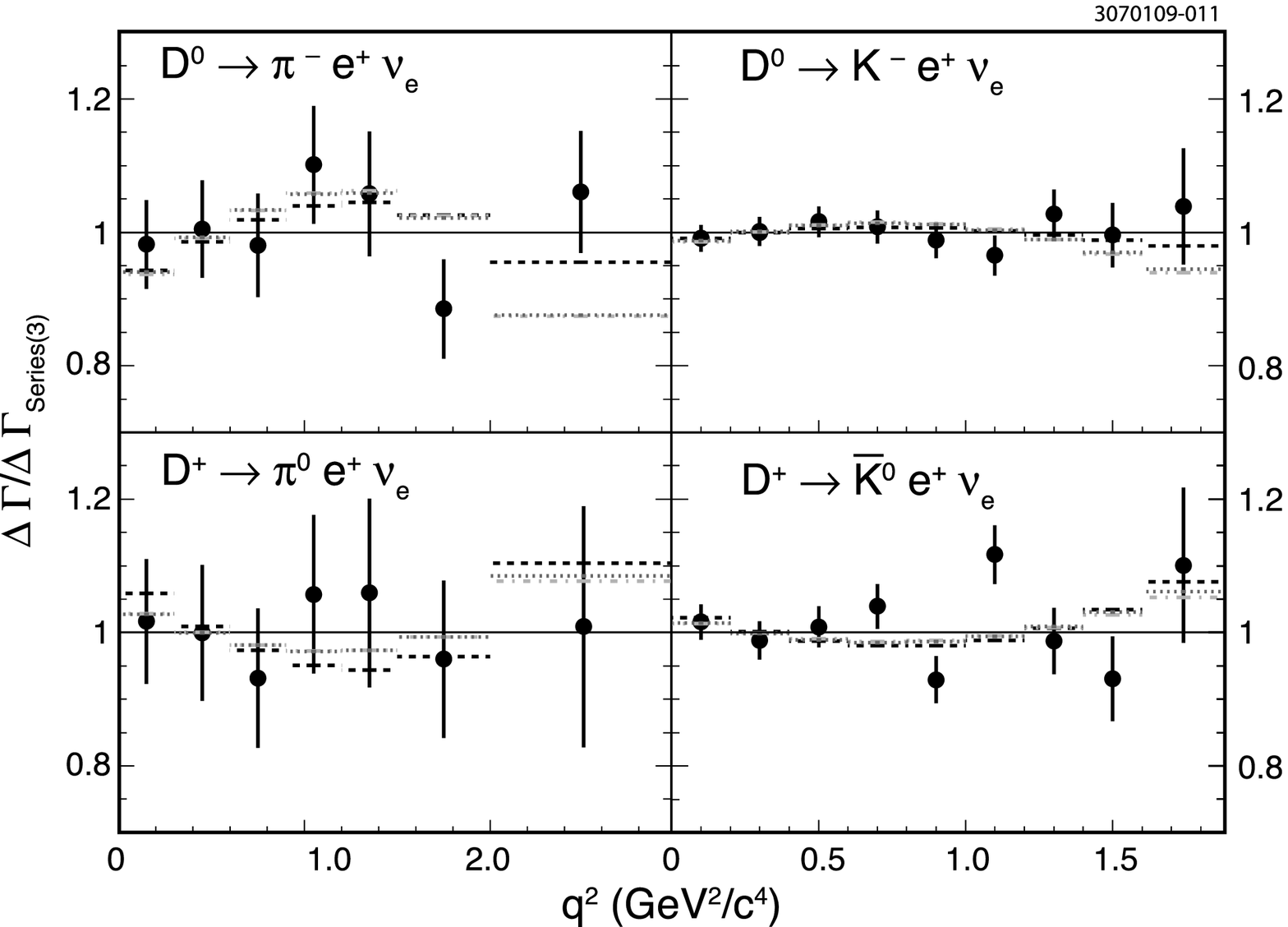}
\caption{Comparison of form factor fits for each semileptonic mode.
The data (squares) and fits to the form factor parameterizations (histograms), including
the simple pole model (long dash),
modified pole model (short dash),
and two parameter series fit (dotted),
 are all normalized using the
three parameter fit result (solid line at unity).
}
\label{fig:dGam_5models}
\end{figure*}

In Fig.~\ref{fig:dGam_5models}, we compare the form factor fits for each of the four semileptonic modes (ISGW2 is excluded).  The partial rates ($\Delta \Gamma$) obtained from each fit have been normalized using those from the three parameter series fits ($\Delta \Gamma_{{\rm Series(3)}}$).  We note that the two-parameter series and modified pole models give nearly indistinguishable results.  Comparing the two and three parameter series formulations, both fits are of reasonable quality.  In all modes but the $\neutralkenu$ (where all parameterizations have slightly large values of $\chi^2$ due to a statistical fluctuation between the fifth and sixth $q^2$ bins), the $\chi^2$ per degree of freedom using a three parameter fit is smaller than that obtained with a two parameter fit.  The strongest evidence for a non-zero value of $a_2$ is in $\chargedpienu$, where $r_2=a_2/a_0$ is slightly more than two standard deviations larger than zero.  Thus, although there are hints of a preference for the three parameter fit, we do not have sufficient statistical evidence to draw strong conclusions on this point.

In general, the quality of the fits is good for all parameterizations; as long as the normalization and at least one shape parameter are allowed to float, all models describe the data well.  We take the three-parameter series fits as our nominal fits, using these to extract $f_+\left(0\right)$, CKM parameters and branching fractions.  We base this decision on the optimized convergence properties of the series expansion as well as the indications discussed above that the data prefer a 3-parameter fit.

Our results rule out the predicted value of $r_{\rm ISGW2}$ and $m_{\rm pole}$ in the ISGW and simple pole models respectively, as have previous studies \cite{Ge:2008yi,Cronin-Hennessey:2008}.   Calculating $1+1/\beta-\delta$, defined in Eq.~(\ref{eq:betadelta}), using the results of the isospin-combined three parameter series fits, we find
\begin{equation}
{1+1/\beta-\delta\left(D\rightarrow\pi^-/\pi^0e^+\nu_e\right)}  = 0.93\pm0.09\pm0.01 ,
\end{equation}
and
\begin{equation}
{1+1/\beta-\delta \left(D\rightarrow K^-/\bar{K}^0e^+\nu_e\right)}  =  0.89\pm0.04\pm0.01 .
\end{equation}
Our values, which are in agreement with~\cite{Ge:2008yi,Cronin-Hennessey:2008} but are more precise, do not support the assumption by the modified pole model that
$1+1/\beta-\delta\approx 2$.

Table~\ref{tab:ff_comp} shows a comparison of our measurements of $f_+(0)$ with other experimental measurements and theoretical predictions.  Our results are taken from the isospin-combined three parameter series expansion fits and include a third uncertainty from the CKM elements, which we take from Particle Data Group fits assuming CKM unitarity~\cite{Amsler:2008zz}.  Our results are in good agreement with previous measurements; they are also consistent with LQCD predictions, although the currently available LQCD results have relatively large uncertainties.

Many previous form factor measurements and predictions have used the modified pole model.  A comparison of our measurements of the shape parameter $\alpha$ is shown in Table~\ref{tab:alpha_comp}.  Other experimental results are generally compatible with our results.  The LQCD results are higher than ours by 2.1 and 2.3 standard deviations for pion and kaon final states, respectively.  We note that variations in $\alpha$ between studies may reflect sensitivities to different regions of $q^2$ coupled with an imperfect parameterization rather than disagreements in the observed form factor distributions.  Our measured form factors in each $q^2$ bin have been compared to LQCD calculations~\cite{Aubin:2004ej}, as shown in Fig.~\ref{fig:isospin}.

\begin{table}
\caption{Comparison of form factor normalization results to previous results.  The third uncertainties show CKM uncertainties where applicable.}
\begin{tabular}{lcc}
\hline\hline
 & $f_+^K(0)$  & $f_+^\pi(0)$ \rule[-1mm]{-1mm}{4.3mm}\\ \hline
LQCD1 \cite{Abada:2002xe}    & 0.66(4)(1) & 0.57(6)(2) \rule[-1mm]{-1mm}{4.3mm}\\
LQCD2 \cite{Aubin:2004ej} & 0.73(3)(7) & 0.64(3)(6) \\
Belle \cite{Widhalm:2006wz} & 0.695(7)(22) & 0.624(20)(30) \\
BABAR \cite{Aubert:2006mc} & 0.727(7)(5)(7) & \\
CLEO-c (281 $\invpb$) \cite{Ge:2008yi} & 0.763(7)(6)(0) & 0.629(22)(7)(3) \\
CLEO-c (this work)  & 0.739(7)(5)(0) & 0.666(19)(4)(3) \\
\hline\hline
\end{tabular}
\label{tab:ff_comp}
\end{table}

The BaBar experiment has reported form factor results using the series expansion for $\chargedkenu$. They find $r_1=-2.5\pm0.2\pm0.2$, and $r_2=0.6\pm6.0\pm5.0$~\cite{Aubert:2006mc}. Our earlier study~\cite{Ge:2008yi,Cronin-Hennessey:2008} measured $r_1$ and $r_2$ for both $\chargedkenu$ and $\chargedpienu$. The results reported here are in agreement with previous results and are the most precise for $\chargedpienu$.

\begin{table}
\caption{Comparison of form factor shape parameter $\alpha$, from fits using the modified pole model, with previous results.}
\begin{tabular}{lcc}
\hline\hline
 & $\alpha^K$  & $\alpha^\pi$ \rule[-1mm]{-1mm}{4.3mm}\\ \hline
LQCD \cite{Aubin:2004ej} & 0.50(4)(7) & 0.44(4)(7) \rule[-1mm]{-1mm}{4.3mm} \\
FOCUS \cite{Link:2004dh} & 0.28(8)(7) &  \\
CLEO III \cite{Huang:2004fra} & 0.36(10)(5) & 0.37(25)(15) \\
Belle \cite{Widhalm:2006wz} & 0.52(8)(6) & 0.10(21)(10) \\
BABAR \cite{Aubert:2006mc} & 0.377(23)(29) & \\
CLEO-c (281 $\invpb$) \cite{Ge:2008yi} & 0.21(5)(2) & 0.16(10)(5) \\
CLEO-c (281 $\invpb$) \cite{Cronin-Hennessey:2008} & 0.21(5)(3) & 0.37(8)(3) \\
CLEO-c (this work)  & 0.30(3)(1) & 0.21(7)(2) \\
\hline\hline
\end{tabular}
\label{tab:alpha_comp}
\end{table}

\subsection{Branching Fraction Results}
Branching fractions are extracted from the three parameter series expansion fit by integrating the optimized fit results over $q^2$.  We find
\begin{equation}
{\mathcal B\left(\chargedpienu\right)}  = (0.288\pm0.008\pm0.003)\% ,
\end{equation}
\begin{equation}
{\mathcal B \left(\chargedkenu\right)}  =  (3.50\pm0.03\pm0.04)\% ,
\end{equation}
\begin{equation}
{\mathcal B \left(\neutralpienu\right)} =  (0.405\pm0.016\pm0.009)\% ,
\end{equation}
and
\begin{equation}
{\mathcal B \left(\neutralkenu\right)} =  (8.83\pm0.10\pm0.20)\% .
\end{equation}
A comparison of these branching fractions with previous measurements is shown in Table~\ref{tab:bf_comp}.  Included in the table are the averaged results of a tagged and an untagged analysis of the initial 281 \nolinebreak $\invpb$ of CLEO-c data; differences between the results reported here and those of previous CLEO-c measurements are within statistical and systematic uncertainties.  We also find that the branching fractions reported here are in excellent agreement with results from other experiments, but are more precise.  This precision arises partially from the ease with which $D$ production can be determined for data collected at the $\psi(3770)$.

\begingroup
\begin{table*}
\caption{Comparison of branching fraction results (\%) in this analysis to previous results.}
\begin{tabular}{lcccc}
\hline\hline
 & $\chargedpienu$ & $\chargedkenu$ & $\neutralpienu$ & $\neutralkenu$ \\ \hline
BES II~\cite{Ablikim:2004ej}        &               & 3.82(40)(27)      &               & 8.71(38)(37) \rule[-1mm]{-1mm}{4.3mm}\\
Belle~\cite{Widhalm:2006wz}          & 0.279(27)(16) & 3.45(10)(19)      &               &              \\
BABAR~\cite{Aubert:2006mc}           &               & 3.522(27)(45)(65) &               &              \\
CLEO-c (281 $\invpb$)~\cite{Ge:2008yi} & 0.304(11)(5)  & 3.60(3)(6)        & 0.378(20)(12) & 8.69(12)(19) \\
CLEO-c (this work)                 & 0.288(8)(3)   & 3.50(3)(4)        & 0.405(16)(9)  & 8.83(10)(20) \\

\hline\hline
\end{tabular}
\label{tab:bf_comp}
\end{table*}
\endgroup

\subsection{Extraction of $|\vcd|$ and $|\vcs|$}
To extract the magnitudes of CKM matrix elements $\left|\vcd\right|$ and $\left|\vcs\right|$, we take the $\left|\vcq\right|\fz$ values from the isospin-combined three parameter series expansion fits and use the LQCD measurements \cite{Aubin:2004ej} $f_+(0)=0.64\pm0.03\pm0.06$ for $D\rightarrow\pi$ transitions and $f_+(0)=0.73\pm0.03\pm0.07$ for $D\rightarrow K$ transitions.  We find
\begin{equation}
\left|\vcd\right| = 0.234\pm0.007\pm0.002\pm0.025
\end{equation}
and
\begin{equation}
\left|\vcs\right| = 0.985\pm0.009\pm0.006\pm0.103,
\end{equation}
where the third uncertainties are from $f_+(0)$.  These are in agreement with those reported by the Particle Data Group (based on the assumption of CKM unitarity)~\cite{Amsler:2008zz}.

\section{Conclusion}
\label{sec:conclusion}
We have described measurements of the $q^2$ dependent partial rates of $\chargedpienu$, $\chargedkenu$, $\neutralpienu$, and $\neutralkenu$.  We have used the partial rates to extract form factor parameters and branching fractions.  Taking input from LQCD, we have reported measurements of $|\vcd|$ and $|\vcs|$. The work here uses the entire CLEO-c sample of $\psi(3770)\rightarrow D\bar{D}$ events and supersedes all previously published CLEO-c studies of $\chargedpienu$, $\chargedkenu$, $\neutralpienu$, and $\neutralkenu$.  Our measurements of branching fractions  and $D\rightarrow\pi^-/\pi^{0}e^+\nu_e$ form factor parameters are the most precise to date.  The results reported here are in agreement with LQCD and will be an incisive test of future calculations.


\section{Acknowledgments}

We gratefully acknowledge the effort of the CESR staff
in providing us with excellent luminosity and running conditions.
D.~Cronin-Hennessy and A.~Ryd thank the A.P.~Sloan Foundation.
This work was supported by the National Science Foundation,
the U.S. Department of Energy,
the Natural Sciences and Engineering Research Council of Canada, and
the U.K. Science and Technology Facilities Council.
We thank A.~Kronfeld for valuable discussions.

\appendix*
\section{Correlation Matrices}
\label{sec:app1}
The statistical correlation matrices, described in Sec.~\ref{sec:signal_yields} and uncorrelated across the semileptonic modes, are shown in Tables~\ref{tab:stat_cov_pi} and~\ref{tab:stat_cov_k}.  The systematic correlation matrices are shown in Tables~\ref{tab:sys_cov_pi} and~\ref{tab:sys_cov_k}.  The diagonal blocks relating systematic uncertainties within a particular mode are constructed as described in Sec.~\ref{sec:systematics}.  To form the matrix elements for the off-diagonal blocks, we have assumed that the uncertainties related to tracking, tag line shapes, fake tags, positron identification, and FSR are fully correlated across semileptonic mode while all other systematic uncertainties are uncorrelated.

\begin{turnpage}
\begingroup
\squeezetable
\begin{table*}
\caption{Statistical correlation matrix for $\chargedpienu$ and $\neutralpienu$ using the standard $q^2$ binning.  $q^2$ increases from left to right and from top to bottom.}
  \label{tab:stat_cov_pi}
  \begin{ruledtabular}
    \begin{tabular}{lddddddd|ddddddd} & \multicolumn{7}{c|}{$\chargedpienu$}&\multicolumn{7}{c}{$\neutralpienu$} \\  & 1.000 & -0.050 & 0.001 & 0.000 & 0.000 & 0.000 & 0.000 & 0.000 & 0.000 & 0.000 & 0.000 & 0.000 & 0.000 & 0.000 \\
 &  & 1.000 & -0.060 & 0.001 & 0.000 & 0.000 & 0.000 & 0.000 & 0.000 & 0.000 & 0.000 & 0.000 & 0.000 & 0.000 \\
 &  &  & 1.000 & -0.060 & 0.001 & 0.000 & 0.000 & 0.000 & 0.000 & 0.000 & 0.000 & 0.000 & 0.000 & 0.000 \\
$\chargedpienu$  &  &  &  & 1.000 & -0.062 & 0.000 & 0.000 & 0.000 & 0.000 & 0.000 & 0.000 & 0.000 & 0.000 & 0.000 \\
 &  &  &  &  & 1.000 & -0.046 & 0.000 & 0.000 & 0.000 & 0.000 & 0.000 & 0.000 & 0.000 & 0.000 \\
 &  &  &  &  &  & 1.000 & -0.030 & 0.000 & 0.000 & 0.000 & 0.000 & 0.000 & 0.000 & 0.000 \\
 &  &  &  &  &  &  & 1.000 & 0.000 & 0.000 & 0.000 & 0.000 & 0.000 & 0.000 & 0.000 \\
\hline
 &  &  &  &  &  &  &  & 1.000 & -0.092 & 0.008 & 0.000 & 0.000 & 0.000 & -0.005 \\
 &  &  &  &  &  &  &  &  & 1.000 & -0.121 & 0.011 & 0.000 & 0.000 & -0.006 \\
 &  &  &  &  &  &  &  &  &  & 1.000 & -0.133 & 0.011 & -0.001 & -0.008 \\
$\neutralpienu$  &  &  &  &  &  &  &  &  &  &  & 1.000 & -0.128 & 0.006 & -0.013 \\
 &  &  &  &  &  &  &  &  &  &  &  & 1.000 & -0.104 & -0.016 \\
 &  &  &  &  &  &  &  &  &  &  &  &  & 1.000 & -0.095 \\
 &  &  &  &  &  &  &  &  &  &  &  &  &  & 1.000 \\
    \end{tabular}
  \end{ruledtabular}
 \end{table*}
\endgroup
\end{turnpage}
\begin{turnpage}
\begingroup
\squeezetable
\begin{table*}
  \label{tab:sys_cov_pi}
\caption{Systematic correlation matrix for $\chargedpienu$ and $\neutralpienu$ using the standard $q^2$ binning.  $q^2$ increases from left to right and from top to bottom.}
  \begin{ruledtabular}
    \begin{tabular}{lccccccc|ccccccc} & \multicolumn{7}{c|}{$\chargedpienu$}&\multicolumn{7}{c}{$\neutralpienu$} \\  & 1.000 & 0.716 & 0.595 & 0.643 & 0.838 & 0.337 & 0.409 & 0.219 & 0.338 & 0.207 & 0.241 & 0.145 & 0.144 & 0.064 \\
 &  & 1.000 & 0.950 & 0.942 & 0.875 & 0.831 & 0.772 & 0.278 & 0.428 & 0.264 & 0.310 & 0.189 & 0.188 & 0.083 \\
 &  &  & 1.000 & 0.975 & 0.864 & 0.906 & 0.826 & 0.247 & 0.379 & 0.235 & 0.277 & 0.169 & 0.169 & 0.074 \\
$\chargedpienu$  &  &  &  & 1.000 & 0.912 & 0.886 & 0.793 & 0.253 & 0.384 & 0.242 & 0.290 & 0.179 & 0.181 & 0.080 \\
 &  &  &  &  & 1.000 & 0.714 & 0.709 & 0.245 & 0.371 & 0.236 & 0.286 & 0.178 & 0.183 & 0.083 \\
 &  &  &  &  &  & 1.000 & 0.862 & 0.212 & 0.318 & 0.205 & 0.253 & 0.161 & 0.170 & 0.083 \\
 &  &  &  &  &  &  & 1.000 & 0.188 & 0.281 & 0.181 & 0.227 & 0.148 & 0.166 & 0.092 \\
\hline
 &  &  &  &  &  &  &  & 1.000 & 0.773 & 0.171 & 0.614 & 0.675 & 0.273 & -0.182 \\
 &  &  &  &  &  &  &  &  & 1.000 & 0.371 & 0.709 & 0.600 & 0.625 & -0.031 \\
 &  &  &  &  &  &  &  &  &  & 1.000 & -0.051 & 0.713 & 0.480 & 0.480 \\
$\neutralpienu$  &  &  &  &  &  &  &  &  &  &  & 1.000 & 0.229 & 0.497 & 0.110 \\
 &  &  &  &  &  &  &  &  &  &  &  & 1.000 & 0.536 & 0.145 \\
 &  &  &  &  &  &  &  &  &  &  &  &  & 1.000 & 0.273 \\
 &  &  &  &  &  &  &  &  &  &  &  &  &  & 1.000 \\
    \end{tabular}
  \end{ruledtabular}
 \end{table*}
\endgroup
\end{turnpage}
\begin{turnpage}
\begingroup
\squeezetable
\begin{table*}
\caption{Statistical correlation matrix for $\chargedkenu$ and $\neutralkenu$ using the standard $q^2$ binning.  $q^2$ increases from left to right and from top to bottom.}
  \label{tab:stat_cov_k}
  \begin{ruledtabular}
    \begin{tabular}{lddddddddd|ddddddddd} & \multicolumn{9}{c|}{$\chargedkenu$}&\multicolumn{9}{c}{$\neutralkenu$} \\  & 1.00 & -0.05 & 0.01 & 0.00 & 0.00 & 0.00 & 0.00 & 0.00 & 0.00 & 0.00 & 0.00 & 0.00 & 0.00 & 0.00 & 0.00 & 0.00 & 0.00 & 0.00 \\
 &  & 1.00 & -0.07 & 0.01 & 0.00 & 0.00 & 0.00 & 0.00 & 0.00 & 0.00 & 0.00 & 0.00 & 0.00 & 0.00 & 0.00 & 0.00 & 0.00 & 0.00 \\
 &  &  & 1.00 & -0.07 & 0.00 & 0.00 & 0.00 & 0.00 & 0.00 & 0.00 & 0.00 & 0.00 & 0.00 & 0.00 & 0.00 & 0.00 & 0.00 & 0.00 \\
 &  &  &  & 1.00 & -0.07 & 0.00 & 0.00 & 0.00 & 0.00 & 0.00 & 0.00 & 0.00 & 0.00 & 0.00 & 0.00 & 0.00 & 0.00 & 0.00 \\
$\chargedkenu$  &  &  &  &  & 1.00 & -0.06 & 0.00 & 0.00 & 0.00 & 0.00 & 0.00 & 0.00 & 0.00 & 0.00 & 0.00 & 0.00 & 0.00 & 0.00 \\
 &  &  &  &  &  & 1.00 & -0.06 & 0.00 & 0.00 & 0.00 & 0.00 & 0.00 & 0.00 & 0.00 & 0.00 & 0.00 & 0.00 & 0.00 \\
 &  &  &  &  &  &  & 1.00 & -0.05 & 0.00 & 0.00 & 0.00 & 0.00 & 0.00 & 0.00 & 0.00 & 0.00 & 0.00 & 0.00 \\
 &  &  &  &  &  &  &  & 1.00 & -0.04 & 0.00 & 0.00 & 0.00 & 0.00 & 0.00 & 0.00 & 0.00 & 0.00 & 0.00 \\
 &  &  &  &  &  &  &  &  & 1.00 & 0.00 & 0.00 & 0.00 & 0.00 & 0.00 & 0.00 & 0.00 & 0.00 & 0.00 \\
\hline
 &  &  &  &  &  &  &  &  &  & 1.00 & -0.05 & 0.01 & 0.00 & 0.00 & 0.00 & 0.00 & 0.00 & 0.00 \\
 &  &  &  &  &  &  &  &  &  &  & 1.00 & -0.07 & 0.01 & 0.00 & 0.00 & 0.00 & 0.00 & 0.00 \\
 &  &  &  &  &  &  &  &  &  &  &  & 1.00 & -0.07 & 0.01 & 0.00 & 0.00 & 0.00 & 0.00 \\
 &  &  &  &  &  &  &  &  &  &  &  &  & 1.00 & -0.07 & 0.00 & 0.00 & 0.00 & 0.00 \\
$\neutralkenu$  &  &  &  &  &  &  &  &  &  &  &  &  &  & 1.00 & -0.07 & 0.00 & 0.00 & 0.00 \\
 &  &  &  &  &  &  &  &  &  &  &  &  &  &  & 1.00 & -0.06 & 0.00 & -0.00 \\
 &  &  &  &  &  &  &  &  &  &  &  &  &  &  &  & 1.00 & -0.06 & 0.00 \\
 &  &  &  &  &  &  &  &  &  &  &  &  &  &  &  &  & 1.00 & -0.05 \\
 &  &  &  &  &  &  &  &  &  &  &  &  &  &  &  &  &  & 1.00 \\
    \end{tabular}
  \end{ruledtabular}
 \end{table*}
\endgroup
\end{turnpage}
\begin{turnpage}
\begingroup
\squeezetable
\begin{table*}
\caption{Systematic correlation matrix for $\chargedkenu$ and $\neutralkenu$ using the standard $q^2$ binning.  $q^2$ increases from left to right and from top to bottom.}
  \label{tab:sys_cov_k}
  \begin{ruledtabular}
    \begin{tabular}{lddddddddd|ddddddddd} & \multicolumn{9}{c|}{$\chargedkenu$}&\multicolumn{9}{c}{$\neutralkenu$} \\  & 1.00 & 0.80 & 0.87 & 0.78 & 0.79 & 0.56 & 0.81 & 0.51 & 0.03 & 0.44 & 0.46 & 0.47 & 0.48 & 0.46 & 0.51 & 0.55 & 0.52 & 0.33 \\
 &  & 1.00 & 0.97 & 0.98 & 0.96 & 0.92 & 0.92 & 0.87 & 0.57 & 0.52 & 0.54 & 0.55 & 0.57 & 0.54 & 0.60 & 0.65 & 0.62 & 0.39 \\
 &  &  & 1.00 & 0.97 & 0.96 & 0.86 & 0.94 & 0.82 & 0.45 & 0.51 & 0.54 & 0.55 & 0.56 & 0.54 & 0.60 & 0.65 & 0.62 & 0.39 \\
 &  &  &  & 1.00 & 0.98 & 0.94 & 0.94 & 0.90 & 0.59 & 0.51 & 0.53 & 0.55 & 0.56 & 0.54 & 0.60 & 0.65 & 0.62 & 0.39 \\
$\chargedkenu$  &  &  &  &  & 1.00 & 0.94 & 0.97 & 0.90 & 0.57 & 0.50 & 0.53 & 0.54 & 0.56 & 0.54 & 0.61 & 0.66 & 0.63 & 0.40 \\
 &  &  &  &  &  & 1.00 & 0.89 & 0.97 & 0.80 & 0.46 & 0.48 & 0.50 & 0.52 & 0.50 & 0.56 & 0.61 & 0.59 & 0.38 \\
 &  &  &  &  &  &  & 1.00 & 0.88 & 0.49 & 0.48 & 0.51 & 0.53 & 0.54 & 0.53 & 0.60 & 0.66 & 0.63 & 0.41 \\
 &  &  &  &  &  &  &  & 1.00 & 0.82 & 0.42 & 0.45 & 0.46 & 0.48 & 0.47 & 0.53 & 0.59 & 0.57 & 0.38 \\
 &  &  &  &  &  &  &  &  & 1.00 & 0.26 & 0.27 & 0.28 & 0.29 & 0.29 & 0.33 & 0.36 & 0.35 & 0.24 \\
\hline
 &  &  &  &  &  &  &  &  &  & 1.00 & 0.94 & 0.96 & 0.98 & 0.75 & 0.96 & 0.77 & 0.75 & 0.66 \\
 &  &  &  &  &  &  &  &  &  &  & 1.00 & 0.99 & 0.97 & 0.93 & 0.88 & 0.91 & 0.91 & 0.81 \\
 &  &  &  &  &  &  &  &  &  &  &  & 1.00 & 0.99 & 0.90 & 0.92 & 0.89 & 0.89 & 0.78 \\
 &  &  &  &  &  &  &  &  &  &  &  &  & 1.00 & 0.84 & 0.96 & 0.85 & 0.84 & 0.73 \\
$\neutralkenu$  &  &  &  &  &  &  &  &  &  &  &  &  &  & 1.00 & 0.70 & 0.97 & 0.98 & 0.88 \\
 &  &  &  &  &  &  &  &  &  &  &  &  &  &  & 1.00 & 0.77 & 0.75 & 0.64 \\
 &  &  &  &  &  &  &  &  &  &  &  &  &  &  &  & 1.00 & 0.99 & 0.91 \\
 &  &  &  &  &  &  &  &  &  &  &  &  &  &  &  &  & 1.00 & 0.91 \\
 &  &  &  &  &  &  &  &  &  &  &  &  &  &  &  &  &  & 1.00 \\
    \end{tabular}
  \end{ruledtabular}
\end{table*}
\endgroup
\end{turnpage}

\clearpage



\end{document}